\documentclass[useAMS,usenatbib]{mn2e}

\usepackage{epsfig,natbib}
\usepackage{myaasmacros,amssymb}
\usepackage{graphicx}
\usepackage{placeins}
\usepackage{times}

\newcommand{\dd}{\mathrm{d}}

\def \cii {{\rm C\,{\sc ii]}~}}
\def \ciis {{\rm C\,{\sc ii}$^{\star}$}}
\def \siii {{\rm Si\,{\sc ii}}}
\def \civ {{\rm C\,{\sc iv}}}
\def \hi {{\,\rm H\,{\sc i}}}
\def \hii {{\,\rm H\,{\sc ii}}}
\def \cm {{\,\rm cm}}
\def \Mpc{{\,\rm Mpc}}
\def \kms{{\,\rm km\textrm{ }s^{-1}}}
\def \kpc{\,\mathrm{kpc}}
\def \pc{{\,\rm pc}}
\def \msol{\mathrm{M}_{\odot}}
\def \himath{\mathrm{HI}}
\def \yr {\mathrm{yr}}

\def \mvir {M_{\mathrm{vir}}}
\def \sdla {\sigma_{\mathrm{DLA}}}

\def\lsim{\mathrel{\rlap{\lower3pt\hbox{$\sim$}}
    \raise1pt\hbox{$<$}}}                
\def\gsim{\mathrel{\rlap{\lower3pt\hbox{$\sim$}}
    \raise1pt\hbox{$>$}}}                

\newcommand{\todo}[1]{{\textbf{TO~DO:  #1 }}}

\newcommand{\aponly}[1]{}
\begin{document}

\pubyear{2008}
\title[Damped Lyman Alpha Systems]{Damped Lyman Alpha Systems in Galaxy Formation Simulations}
\author[A. Pontzen et al.]{Andrew Pontzen$^{1}$\thanks{Email:
    apontzen@ast.cam.ac.uk}, Fabio Governato$^{2}$, Max Pettini$^{1}$, C.M. Booth$^{3,4}$,  Greg Stinson$^{2,5}$, \newauthor
James Wadsley$^{5}$,  Alyson Brooks$^{2}$, Thomas Quinn$^{2}$,  Martin Haehnelt$^1$ \\
  $^{1}$Institute of Astronomy, Madingley Road, Cambridge CB3 0HA, UK \\
  $^{2}$Astronomy Department, Box 351580, University of Washington,
  Seattle, WA 98195, USA \\
  $^{3}$Department of Physics, Institute for Computational Cosmology, University of Durham, South Road, Durham, UK \\
  $^{4}$Sterrewacht Leiden, University of Leiden, P.O. Box 9513, 2300 RA Leiden, The Netherlands \\
  $^{5}$Department of Physics and Astronomy, McMaster University, Hamilton, ON L8S 4M1, Canada }

\date{Accepted 2008 July 30. Received 2008 July 30; in original form 2008 April 19}
\maketitle

\begin{abstract}
  We investigate the population of $z=3$ damped Lyman alpha systems
  (DLAs) in a recent series of high resolution galaxy formation
  simulations. The simulations are of interest because they form at
  $z=0$ some of the most realistic disk galaxies to date. No free
  parameters are available in our study: the simulation parameters
  have been fixed by physical and $z=0$ observational constraints, and
  thus our work provides a genuine consistency test. The precise role
  of DLAs in galaxy formation remains in debate, but they provide a
  number of strong constraints on the nature of our simulated bound
  systems at $z=3$ because of their coupled information on neutral
  \hi~densities, kinematics, metallicity and estimates of star
  formation activity.

  Our results, without any parameter-tuning, closely match the
  observed incidence rate and column density distributions of
  DLAs. Our simulations are the first to reproduce the distribution of
  metallicities (with a median of $Z_{\mathrm{DLA}}\simeq
  Z_{\odot}/20$) without invoking observationally unsupported
  mechanisms such as significant dust biasing. This is especially
  encouraging given that these simulations have previously been shown
  to have a realistic $0<z<2$ stellar mass-metallicity
  relation. Additionally, we see a strong positive correlation between
  sightline metallicity and low-ion velocity width, the normalization
  and slope of which comes close to matching recent observational
  results. However, we somewhat underestimate the number of observed
  high velocity width systems; the severity of this disagreement is
  comparable to other recent DLA-focused studies.

  DLAs in our simulations are predominantly associated with dark
  matter haloes with virial masses in the range
  $10^9<M_{\mathrm{vir}}/\msol<10^{11}$. We are able to probe DLAs at
  high resolution, irrespective of their masses, by using a range of
  simulations of differing volumes. The fully constrained feedback
  prescription in use causes the majority of DLA haloes to form stars
  at a very low rate, accounting for the low metallicities. It is also
  responsible for the mass-metallicity relation which appears
  essential for reproducing the velocity-metallicity correlation. By
  $z=0$ the majority of the $z=3$ neutral gas forming the DLAs has
  been converted into stars, in agreement with rough physical
  expectations.
\end{abstract}


\section{Introduction}\label{sec:introduction}

One of the most difficult and most important questions to ask of any
cosmological simulation is whether, even if it matches some known
properties of the observed Universe, the route via which it obtained
those results is physically meaningful. It is tempting to argue that,
with the degree of parameter-tuning available to the modern simulator
(stemming from our inability to maintain a sufficient dynamic range,
uncertainty in gas physics and in particular star formation and
feedback prescriptions), attempts to match a small number of observed
properties can succeed without representing a physical route to that
success.  A sensible test of any suite of simulations, therefore, is
to scrutinize its predictions for observed relations which were not
considered in the process of planning those simulations. Success or
failure of the simulation to match such relations cannot be equated to
success or failure of the simulation and its predictions as a whole,
but it can lend weight in either direction.

In this paper we will apply such an approach to the series of galaxy
formation simulations most recently described in
\cite{2007MNRAS.374.1479G}, \cite{2007ApJ...655L..17B} and
\cite{2008arXiv0801.1707G} (henceforth G07, B07 \& G08 respectively). The
$z=0$ outputs of these simulations contain more realistic disc
galaxies than have previously been achieved. In particular, the
simulated galaxies form rotationally supported disks falling on the
$z=0$ Tully-Fisher and baryonic Tully-Fisher relations, have a
distribution of satellites compatible with local observations (G07)
and have reasonable stellar mass-metallicity relations
(B07).

It should be noted, however, that in common with other galaxy
formation simulations, the mass in the bulge component of the G07
galaxies is overestimated \citep[see also][]{2001ApJ...554..114E}. The
problem is essentially one of angular momentum loss; it seems that to
prevent this, both high numerical resolution
\citep{2007MNRAS.375...53K} and better models of feedback from
supernova explosions are important elements (G08).  The exaggerated
bulges cause rotation curves to decline unrealistically over a few
disk scale lengths to $\sim 75 \%$ of their peak value. However, these
problems appear to be shrinking in magnitude as resolution increases
-- and, regardless, the simulations under consideration form galaxies
at $z=0$ which are as realistic as current computational and modeling
power will allow, so that a detailed study of their properties during
formation is justified.

In this paper, we will investigate at $z=3$ the predominantly neutral
gas which gives rise to damped Lyman alpha systems (DLAs). These are
systems with column densities of \hi~ in excess of $2 \times 10^{20}
\cm^{-2}$, seen in absorption against more distant luminous sources
(generally quasars); for a recent review see
\cite{2005ARA&A..43..861W}. The particular limit is historical,
corresponding to the column densities expected if the Milky Way were
to be viewed face on \citep{1986ApJS...61..249W}, but simple physical
arguments suggest it makes a convenient distinction between trace
\hi~in the ionised intergalactic medium (IGM, below the limit) and
clouds which are predominantly composed of \hi~\cite[above the limit,
see][]{2005ARA&A..43..861W}. The latter clouds must absorb, in an
outer layer, the majority of incident photons capable of ionising
hydrogen ($h \nu > 13.6\, \mathrm{eV}$) and are therefore termed
``self-shielding''.

The existence of an intergalactic ultraviolet (UV) field arising from
the cumulative effect of external galaxies and quasars
\citep[e.g.][]{1996ApJ...461...20H} plays many roles in understanding
the state of these clouds -- not only does it affect the ionisation
levels in the optically thin transition regions, but it also
contributes substantially to the heating budget via photo-ejection of
electrons from atoms. Consequently, gas cannot cool to form neutral
clouds without first collapsing in the presence of a gravitational
mass with virial velocity of tens of $\kms$
\citep{1986MNRAS.218P..25R,1996MNRAS.278L..49Q}, suggesting such
clouds are associated with dark matter haloes and hence
protogalaxies. This rough physical argument is verified by previous
simulations (many of which are listed in Table
\ref{tab:previous-simulations}), although in the simulations of
\cite{Razoumov:2005bh} and \cite{2007arXiv0710.4137R} DLAs often
bridge multiple haloes, extending into the intervening IGM. (We discuss
this issue further in Section \ref{sec:no-interg-dlas}.)

\begin{table*}
\begin{tabular}{llllll}
  \hline
  Reference(s) & Type & SF & Ionization/RT & Max Vol$^{(1)}$ & Gas Res$^{(2)}$ \\
  \hline
  \cite{1996ApJ...457L..57K} & SPH & None & Plane Correction$^{(3)}$ & $(22 \Mpc)^3$ & $10^{8.2} \msol$ \\  
  \cite{1997ApJ...484...31G} & SPH  & None  & Plane Correction$^{(3)}$ & $(22 \Mpc)^3$ & $10^{8.2} \msol$ \\
  \cite{1997ApJ...486...42G} \\
  \cite{1998ApJ...495..647H} & SPH & None & Den. Cut$^{(4)}$ & N/A$^{(5)}$ & $10^{6.7} \msol$ \\
  \cite{2001ApJ...559..131G} & SPH & Yes, weak FB$^{(6)}$  & Plane Correction$^{(3)}$ & $(17 \Mpc)^3$ & $10^{8.2} \msol$ \\
  \cite{2003ApJ...598..741C} & Eulerian & Yes, with FB$^{(6)}$ & Hybrid$^{(7)}$ & $(36 \Mpc)^3$ & $11 \kpc$ \\
  \cite{2004MNRAS.348..421N} & SPH & Multiphase/GW$^{(8)}$ & Eq. Thin/MP$^{(8)}$ & $(34 \Mpc)^3$ & $10^{4.6} \msol$\\
  \cite{2004MNRAS.348..435N} \\
  \cite{Razoumov:2005bh} & Adpt Eulerian$^{(9)}$ & None & Non-Eq. Live RT/post-processor$^{(10)}$ & $(8 \Mpc)^3$ & $0.1 \kpc$   \\
  \cite{Nagamine:2005ud} & SPH & Multiphase/GW$^{(8)}$ & Eq. Thin/MP$^{(8)}$ & $(14 \Mpc)^3$ & $10^{5.0} \msol$ \\
  \cite{2007arXiv0710.4137R} & Adpt Eulerian$^{(9)}$ & Basic & Non-Eq. Thin/post-processor$^{(10)}$ & $(45 \Mpc)^3$ & $0.09 \kpc$  \\
  \hline
  This work & SPH & Yes, with FB$^{(6, 11)}$ & Eq. Thin/RT post-processor$^{(11)}$ & $(25 \Mpc)^{3}$ & $10^{4.0} \msol$ \\
  \hline
\end{tabular}
\caption{Selected previous simulations of DLAs. Notes on methods: $^{(1)}$The largest volume simulated for the
  study, in comoving units. $^{(2)}$The best gas resolution achieved in the study, which may not have been achieved in the largest volume. For SPH
  (Lagrangian) simulations we give the smallest particle mass; for Eulerian simulations we give the finest grid resolution (in physical units at $z=3$). $^{(3)}$UV background in optically thin limit; sightlines post-processed using plane parallel radiative transfer and ionisation equilibrium.
  $^{(4)}$UVB optically thin, but in post-processing all gas particles assumed fully neutral for number densities $n>10^{-2} \cm^{-3}$. $^{(5)}$This study used a re-sampling technique to study
  the high resolution dynamics of a limited number of haloes and did not collect cosmological statistics.$^{(6)}$FB = Feedback, i.e. the
  deposition of energy into the ISM due to supernova explosions. By ``Weak'' FB is meant thermal
injection only, which is generally recognized to be insufficient (see Section \ref{sec:feedback}) $^{(7)}$The optical depth for each cell was calculated and used to approximate a local attenuation to the UVB. $^{(8)}$The multiphase method keeps track of the fraction
  of a gas particle in a cold cloud phase in pressure equilibrium with the warm medium; the cold clouds are assumed to be
  fully self shielded while the ambient medium is regarded as optically thin. Phenomenological galactic winds (GW) are added, causing
  bulk outflows from all haloes. $^{(9)}$Adaptive Eulerian: the grids which keep
  track of gas properties are automatically refined as regions collapse on small scales. $^{(10)}$A comprehensive treatment of radiative transfer
  was used by this paper, using eight angular elements in the live simulation and 192 in a post-processing stage. $^{(11)}$See Sections \ref{sec:simulations} and \ref{sec:self-shielding}. 
}\label{tab:previous-simulations}
\end{table*}

Irrespective of their physical nature, it is simple to show directly
that DLAs contain the majority of \hi~over all redshifts $z>0$
\citep[e.g.][]{1987ApJ...321...49T}, which suggests they have an
important role to play in the global star formation
history. Curiously, however, a wide range of diagnostics provide
compelling evidence that the star formation rates in typical DLAs are
small ($\lsim 1 \msol \yr^{-1}$). These include the low characteristic
metallicities and dust depletions \cite[for a review
see][]{2006astro.ph..3066P}, the extreme rarity of detectable optical
counterparts and the faintness of Ly$\alpha$ emission \citep[a large
number of studies are summarised in][]{2005ARA&A..43..861W}. In DLAs
with detectable molecular hydrogen (H$_2$), the local UV can be
estimated from pumping into high energy rotational levels
\cite[e.g.][]{2005MNRAS.362..549S}, results suggesting star formation
rates comparable to the present day Milky Way ($\sim 1\, \msol\,
\yr^{-1}$). However, since few DLAs are associated with detectable
H$_2$ absorption, these measurements are not necessarily
representative of the wider population. The estimated cooling rates
from the recently developed \cii technique \citep{2003ApJ...593..215W}
lead to star formation rate density estimates of $\sim 10^{-2} \msol
\yr^{-1} \kpc^{-2}$, although the exact interpretation of these
results is complicated by various assumptions in the method and the
unknown area of a typical DLA system.

A further constraint on the hosts of DLAs is given by kinematic
information encoded in unsaturated metal absorption lines. Both the
neutral gas (traced by low-ion transitions such as \siii$\lambda
1808$) and any surrounding ionised gas (traced by high-ion transitions
such as \civ$\lambda 1548$) may be probed.  The exact relationship
between the high- and low-ion regions is not entirely certain
\cite[e.g.][ and references therein]{2007A&A...465..171F}.
Emphasis in our work, and most other studies, is placed on the low-ion
profiles since these presumably reflect the kinematics of the gas
giving rise to the DLA itself. 

The earliest systematical survey of DLA low-ion velocity profiles was
conducted by \cite{1997ApJ...487...73P}, who suggested that the
observed kinematics could arise from a population of thick cold
rotating disks with a distribution of rotational velocities similar to
that observed in local disk galaxies.  This is at odds with physical
intuition and the prevailing Cold Dark Matter (CDM) hierarchical
cosmogony in the sense that it requires the halo mass function, and
hence power spectrum of fluctuations, to remain almost unchanged over
$10$ Gyr between $z=3$ and $z=0$. A view more compatible with the
standard model seems desirable. \cite{1998ApJ...495..647H} were quick
to apply numerical simulations of structure formation to show that a
fiducial population of $z=3$ CDM haloes were not incapable of producing
velocity profiles similar to those observed -- however, some details
have proved more problematic as we describe below.

One way of quantifying the simplest property of the kinematics is to
assign to each DLA a ``velocity width'' $v_{90\%}$, roughly measuring
the Doppler broadening of any unsaturated low-ion transition (see
Section \ref{sec:sightline-projection} for a more precise
definition). The velocity width may be presumed to give an indication
of the underlying virial velocity of the system responsible, although
there is no guarantee that a particular sightline will sample the
entire range of the velocity dispersion within the system; the
simulations of \cite{1998ApJ...495..647H} suggested that $v_{90\%}
\sim 0.6 v_{\mathrm{vir}}$ with a large scatter.

Simulations such as those by
\cite{1996ApJ...457L..57K,1997ApJ...484...31G,2001ApJ...559..131G,2004MNRAS.348..421N}
focused instead on the cross-sectional size of haloes as DLAs. Taken
with a halo mass function (which throughout this work, we will produce
for a $\Lambda$CDM ``concordance'' scenario) such trends can be used
to predict the overall rate of occurrence of DLA systems -- a
prediction which, in most simulations, agrees roughly with
observations. Including the results of \cite{1998ApJ...495..647H}, one
can further attempt to predict the relative proportions of systems of
differing velocity width. In general, simulations underestimate the
incidence of high velocity widths; this can be traced to the majority
of the cross-section being assembled from low mass ($\sim 10^9 \msol$)
haloes.  Similar difficulties are also encountered in varying degrees
by semi-analytic models of DLAs. These are, of course, not suited to
producing the details of the line widths but they can help to pin down
which physical considerations affect them \citep[see
e.g.][]{1996MNRAS.281..475K,2001MNRAS.326.1475M,2006MNRAS.371.1519J}.

Overall, the extent of the implications of the velocity mismatch is
somewhat unclear. Some authors have argued that a fundamental
difficulty with the $\Lambda$CDM scenario has been uncovered
\citep{2001ApJ...560L..33P}. However, numerical modeling of the DLA
population is intrinsically troublesome. The dynamic range of
processes involved in making the final population is tremendous, with
the relevant scales ranging from cosmological to stellar. Therefore we
suggest that failure to match -- or, indeed, success in matching --
specific observations should be interpreted conservatively. We now
describe some specific uncertainties in understanding the simulated
DLA population, and outline how these are dealt with in our work. For
comparison, Table \ref{tab:previous-simulations} gives details of
previous simulations and their approach to these problems.

\begin{enumerate}
\item {\it Star Formation.} Although \cite{1997ApJ...484...31G} argue
  that star formation at $z>2$ has little impact on the column
  densities of individual systems, it is not a priori clear how the
  kinematics and cross-section are affected by the supernovae
  feedback. Investigations in this direction have been made by
  \cite{2004MNRAS.348..421N,2004MNRAS.348..435N}, using a
  phenomenological galactic wind prescription. In the present paper,
  our star formation is fully prescribed by physical models and $z=0$
  observations, leaving no parametric freedom; see Section
  \ref{sec:simulations}.
\item {\it Self-Shielding.} DLAs are known to contain gas which is
  largely self-shielding (see above); thus the fiducial ``uniform UV
  background'' which is used in many cosmological simulations proves
  inadequate. We use a simple radiative transfer post-processor to
  correct for this (Section \ref{sec:self-shielding}). We assess both
  the algorithm's reliability and the severity of neglecting radiative
  transfer in the live simulation in Section
  \ref{sec:radiative-transfer}.

\item {\it Cosmological Sampling.} The total DLA cross-section is
  thought to be evenly spread through many orders of magnitude of
  parent halo mass \citep{1997ApJ...484...31G,1997ApJ...486...42G};
  observations of low redshift DLAs appear to confirm such a view
  \citep{2005MNRAS.364.1467Z}. Presumably one must allow a sufficient
  volume in a simulation to study a number of protocluster regions as
  well as maintain sufficient resolution to resolve low mass dwarf
  galaxies. Techniques involving extrapolation into unresolved
  regimes \cite[such as fitting functional forms to the DLA
  cross-section as a function of halo
  mass: ][]{1997ApJ...484...31G,1997ApJ...486...42G} provide useful
  signposts but naturally must be regarded with caution. Our
  simulations, and those by \cite{Razoumov:2005bh} and
  \cite{2004MNRAS.348..421N} resolve all haloes of relevance only by
  separately simulating a number of boxes of varying size; thus a way
  of combining results from the various boxes must be found. In our
  case, this involves using the halo mass function to re-weight
  sightlines in calculating cosmological properties (Section
  \ref{sec:cosm-conv}).
\end{enumerate}
The remainder of this paper is structured as follows. In Sections
\ref{sec:simulations} and \ref{sec:pipeline} we describe respectively
the series of simulations in use for our study and our methods for
extracting DLA sightlines and producing quantities representative of a
cosmological ensemble. We give results for these quantities and their
underlying relations in Section \ref{sec:results}. Section
\ref{sec:consistency-checks} describes a number of consistency checks
and runs with altered parameters which shed further light on the
origin of some of our relations. Finally, we summarise and discuss
future work in Section \ref{sec:discussion}. Two appendices contain
technical detail for completeness.

We adopt the following conventions. Except where specified, all quoted
measurements are given in physical units; where cosmological
parameters enter calculations, these are based on the standard
cosmology used for the simulations: $\Omega_M=0.30$, $\Omega_b=0.044$, 
$\Omega_{\Lambda}=0.70$, $\sigma_8=0.90$, $h=H_0/(100 \kms) = 0.70$,
$n_s=1$. We briefly investigated the effect of incorporating the
favoured parameters from the fifth year WMAP results
\citep{2008arXiv0803.0586D}, but the overall differences are expected
to be minor (Section \ref{sec:cosmological-model}).

\section{Simulations}\label{sec:simulations}

\begin{table}
\begin{center}
\begin{tabular}{lllll}
\hline
Tag & $\langle M_{\mathrm{p,gas}} \rangle$ & $ \langle M_{\mathrm{p,DM}} \rangle$ & $\epsilon/\mathrm{kpc}$ &  Usable Vol (comoving) \\
\hline
Dwf & $10^{4.0} \mathrm{M}_{\odot}$ & $10^{5.0} \mathrm{M}_{\odot}$ & $0.15$ & $5\, \Mpc^3$ \\ 
MW & $10^{5.2}$ & $10^{6.2}$ & $0.31$ & $50\, \Mpc^3$ \\
Large & $10^{6.1}$ & $10^{7.0}$ & $0.53$ &  $600\, \Mpc^3 $ \\
Cosmo & $10^{6.7}$ & $10^{7.6}$ & $1.00$ &  $15625 \Mpc^3$  \\
\hline
\end{tabular}
\end{center}
\caption{The simulations used in this work. The first column is the
  tag which we use to refer to each simulation. For all except ``Cosmo'', a 
  subsample of the full box is simulated in high
  resolution; no results are taken from outside this region.  The second
  and third columns refer respectively to the mean gas and dark matter
  particles within the region, the fourth to the gravitational softening length 
  (in physical units) and the final column gives the comoving volume
  of the region. The separate
  boxes are generated from entirely different sets of initial conditions;
  the ``Dwf'' and ``MW''
  simulations are designed to form respectively a dwarf and Milky Way type galaxy
  at $z=0$ while the ``Large'' and ``Cosmo'' boxes are more statistically representative. 
}\label{tab:simulations}
\end{table}

The simulations are successors to the galaxy formation simulations
described in G07, with higher resolution and a more physical star
formation feedback prescription as described by
\cite{2006astro.ph..2350S} (S06, see also below). They are computed
using the SPH code {\sc Gasoline} \citep{2004NewA....9..137W}, and
include gas cooling and the effects of a uniform ultraviolet (UV)
background \citep[following][]{1996ApJ...461...20H} in an optically
thin approximation. We later perform post-processing to account for
self-shielding effects (Section \ref{sec:self-shielding}). We also ran
test simulations which included an approximate treatment of
self-shielding within the live simulation (Section
\ref{sec:radiative-transfer}).

The SPH smoothing lengths are defined adaptively to use the 32 nearest
neighbours in all averaging calculations, except that a minimum
smoothing length of $0.1$ times the gravitational softening
($\epsilon$) is imposed.  The star formation and feedback recipe (S06)
is based on the algorithms originally described by
\cite{1992ApJ...391..502K}. In short, gas particles can only form
stars if $T<30\,000\,\mathrm{K}$, $n_{\mathrm{gas}}>0.1\cm^{-3}$ and
the local hydrodynamic flow is converging. The rate of star formation
in such regions is assumed to follow the \cite{1959ApJ...129..243S}
law ($\dot{\rho}_{\mathrm{star}} \propto \rho_{\mathrm{gas}}^{n}$)
with index $n=3/2$. This is a rather natural choice of $n$, since it
implies the cold gas is turned into stars over some multiple
$1/c_{\star}$ of the local dynamical timescale:
\begin{equation}
\frac{1}{\rho_{\mathrm{gas}}}\frac{\dd \rho_{\mathrm{star}}}{\dd t} = c_{\star} \sqrt{G \rho_{\mathrm{gas}}} \textbf{.} \label{eq:star-form-rate}
\end{equation}

Following the evolution of each star particle consistently with the
\cite{1993MNRAS.262..545K} IMF, a fixed fraction
$\epsilon_{\mathrm{SN}}$ of the supernova energy created at each
timestep is deposited thermally into the surrounding gas particles. To
emulate the physics of the processes responsible for distributing this
energy to the gas, radiative cooling is disabled in particles within a
radius which must be determined. Traditionally this can involve
further parameters, but the S06 method obviates the need for these by
modeling the physical processes according to a prescription based on
blast wave models
\citep[e.g.][]{1974ApJ...188..501C,1988RvMP...60....1O}: this sets the
radius of the local ISM affected as a function of the local density
and temperature. The free parameters, consisting of the constant of
proportionality in the Schmidt law ($c_{\star}$) and the efficiency of
supernova energy deposition into the ISM ($\epsilon_{\mathrm{SN}}$)
are tuned such that isolated galaxy models match the
\cite{1998ApJ...498..541K} law and cold gas fractions observed in
present day disk galaxies.  This leaves no free parameters in the
context of this study (setting $c_{\star}=0.05$ and
$\epsilon_{\mathrm{SN}}=4 \times 10^{50} \, \mathrm{ergs}$ -- see
\citeauthor{2006astro.ph..2350S} 2006) but produces galaxies which
satisfy a wide range of observational constraints (see Introduction).
As part of the supernova algorithm, metals are also deposited into the
ISM (see Section \ref{sec:metallicity} for more details), but note
that our simulations do not yet contain any contribution to radiative
cooling from metals (which is expected to be physically dominant below
temperatures $T \sim 10^4 \,\mathrm{K}$).

As in G07 \& G08, the prescriptions are implemented in a fully
cosmological context, based on a fiducial model with parameters given
at the end of Section \ref{sec:introduction}. The majority of our
results are derived from four simulations (Table
\ref{tab:simulations}). Three of these take advantage of the ``volume
renormalization'' technique, in which progressively higher resolution
is employed towards the central galaxy of interest
\citep{1993ApJ...412..455K}. The ``Dwf'' and ``MW'' simulations
simulate the same region as the DWF1 and MW1 runs in G07 and G08,
forming at $z=0$ a dwarf and Milky Way type galaxy respectively.  The
third simulation, ``Large'', is a compromise between high resolution
and volume, while the fourth simulation, ``Cosmo'', is of a full $(25
\Mpc)^3$ box populated with gas at uniform resolution. The smoothing
length is constant in physical units for $z<8$, and evolves comovingly
for $z>8$; its final fixed value for each simulation is given in Table
\ref{tab:simulations}. Otherwise, the physics in each of these runs is
identical.

\section{Processing Pipeline}\label{sec:pipeline}
\begin{figure}
\begin{center}
\includegraphics[width=0.49\textwidth]{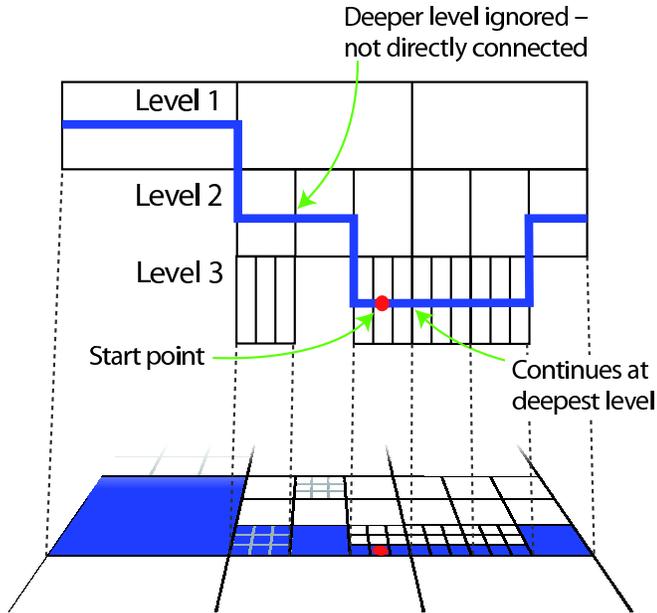}
\end{center}
\caption{A two-dimensional representation of our three-dimensional
  radiative transfer scheme, which involves building a nested grid and
  integrating optical depths from a cell along that grid to the edges
  of the box. A combination of local fine spatial resolution and speed
  is achieved by traversing the grid at the lowest available level
  until this level is closed. At such a point, the algorithm jumps to a
  higher level in the grid. Subsequently, when lower levels become
  available they are ignored since these high density regions are not
  directly connected to the original high density region. This is
  primarily a computational simplification, but it also (in an
  admittedly ad hoc fashion) prevents long distance unphysical
  shadowing which would otherwise arise from our limited angular
  resolution -- see text for details.}\label{fig:rt-diag}
\end{figure}

The processing pipeline is based on {\sc SimAn}\footnote{\tt
  www.ast.cam.ac.uk/\~{}app26/siman/ }, an object-oriented
C++/Python-based environment for analyzing simulations of arbitrary
file format with support for {\sc OpenGL} real-time
visualization. Such a framework allows us to understand our data
rapidly, especially since it includes support for stereoscopic glasses
-- giving a true 3D rendering of the data. It is written in such a way
as to allow analysis of data from an interactive python session,
hiding a large number of details from the scientific user.

\subsection{Self-Shielding}\label{sec:self-shielding}

We employ a basic radiative transfer post-processor to assess the
effect of self-shielding. The simulation is divided into nested grids
so that the lowest level cells contain close to 32 particles, to
correspond with the SPH scheme of G07. We use a cosmic ultraviolet
background (UVB) following a revised version of
\cite{1996ApJ...461...20H} (Haardt 2006, private communication); this
is initially assumed to be present throughout the simulation volume.

In each cell, using the radiation field described above, the
ionization state of the hydrogen and helium gas is determined using an
equilibrium solver algorithm based on \cite{1996ApJS..105...19K}. This
requires the temperature and density of each SPH particle, derived
directly from the simulation output. We keep the temperature, not the
internal energy, fixed during this process. While not ideal, it is
necessary to make some such assumption to determine the thermal state
in post-processing. As a test we verified that keeping instead the
internal energy fixed makes little noticeable difference to our
results.  On the other hand, shielding can significantly drop the
heating rate, so that an assessment of the dynamical effect is
essential; we find it to be minor in terms of our final
results, although uncertainties remain (see Sections
\ref{sec:where-is-siii} and \ref{sec:no-interg-dlas}).

With the ionisation state determined, the attenuation of the UV field
due to the \hi, He\,{\sc i} and He\,{\sc ii} ions is then
calculated. This translates into an optical depth (as a function of
frequency) for the cell. In subsequent iterations, the attenuation of
the incoming UVB radiation is calculated by integrating the cell
optical depths along the six directions defined by the orientation of
the grid to the edge of the box. Because the grid is refined
adaptively, this process is somewhat complicated by the need to
``level-jump'', i.e. move to higher levels of the nested structure
when the lower levels run out. Note that the grid walk stays at the
lowest level in any directly connected region of the origin cell
(including when crossing boundaries of higher levels). As it moves out
of the high resolution region, higher levels are selected as
appropriate. The walk does not re-descend, even if lower levels again
become available, instead using the averaged properties of regions
defined by the current level. This is primarily a matter of
computational speed; however, it has the side-effect of preventing
long distance shadowing which can arise in low angular-resolution
radiative transfer mechanisms. Admittedly we have not formulated this
in terms of any limiting procedure yielding a well-defined physical
calculation, but heuristically the averaging over larger solid angles
as the integration proceeds away from the target cell is quite
correct. We have illustrated this situation in
Figure~\ref{fig:rt-diag}.

This entire process is iterated over the full simulation, convergence
being assessed by changes in the UVB field and ionization state
between steps. We found that the system defined above had some
oscillatory behaviour on its approach to convergence, which led us to
introduce a damping term which in effect averages the optical depth
between iterations. This results in faster convergence, but does not
essentially alter the scheme described.

Our self-shielding process is crude when compared to recent radiative
transfer codes \citep[for a comparison see ][]{2006astro.ph..3199I};
however, we do not believe this makes our results obsolete: see
Section \ref{sec:radiative-transfer}. A further complication must be
considered, which is the failure of the scheme as described to account
for any dynamic effects of the changing ionisation fraction and
heating rates. We have assessed the severity of this problem using a
local attenuation approximation: see Section
\ref{sec:radiative-transfer}.

Figure \ref{fig:projected-hi} shows a $z=3$ map of the neutral
hydrogen in a $200 \kpc$ (physical) cube centred on the major
progenitor of a Milky Way type galaxy in our ``MW'' box. It is
coloured such that DLAs appear in red and Lyman limit systems appear
in green/yellow. The locations and virial radii (see Section
\ref{sec:dla-properties-haloes}, below) of dark matter haloes exceeding
$5 \times 10^8 \msol$ in virial mass are overplotted.

\subsection{Calculation of Halo and Sightline Properties}\label{sec:dla-properties-haloes}\label{sec:sightline-projection}

We start by locating individual dark matter haloes in the simulation
using the grid-based code AHF\footnote{\tt
  http://www.aip.de/People/aknebe/AMIGA}
\citep{2001MNRAS.325..845K,2004MNRAS.351..399G}. We then mark a
spherical region of radius $r_{\mathrm{vir}}$ (the radius within which
the density exceeds the mean density of the simulation at the given
redshift by a factor 178, by analogy with spherical top-hat collapse
models) as belonging to the halo of mass $M_{\mathrm{vir}}$.  We also
record the masses in gas, stars and \hi~ within the virial radius.
The virial velocity is defined as usual, $v_{\mathrm{vir}}^2=
GM_{\mathrm{vir}}/r_{\mathrm{vir}}$.

In the resampled simulations, we immediately discard any haloes which
have been contaminated by low resolution particles from the outer
regions. We also discard any haloes with fewer than 200 gas particles
or 1\,000 dark matter particles. We determined this limit empirically
by examining the point at which halo properties diverged in
simulations at different resolutions -- for more details, see Section
\ref{sec:resolution}.

\begin{figure*}
\begin{center}
\includegraphics[width=0.98\textwidth]{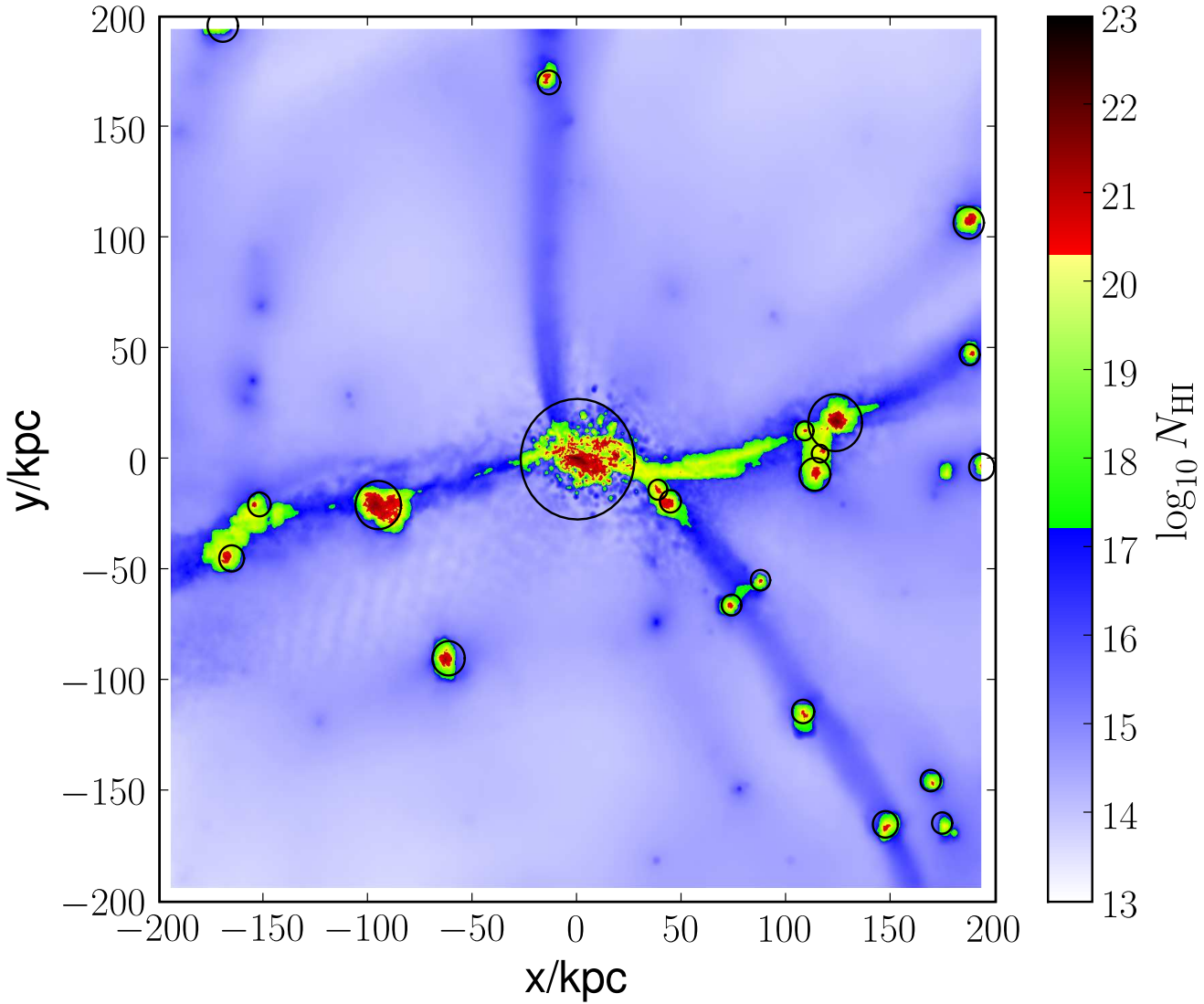}
\end{center}
\caption{The $z=3$ neutral column density of \hi~ in a $400 \kpc$ cube
  centred on the major progenitor to a $z=0$ Milky Way type galaxy
  (box MW). The colours are such that DLAs ($\log_{10}
  N_{\himath}/\cm^{-2} >20.3$) appear in dark red and Lyman limit systems
  ($20.3 > \log_{10} N_{\himath}/\cm^{-2}>17.2$) appear in green and
  yellow. The circles indicate the projected positions and virial
  radii of all dark matter haloes with $M>5 \times 10^{8} \msol$.  All
  units are physical. A stereoscopic version and an animated version
  of this plot are available at {\tt
    www.ast.cam.ac.uk/\~{}app26/}.}\label{fig:projected-hi}
\end{figure*} 

\begin{figure*}
\vspace{1.0cm}
\includegraphics[width=0.48\textwidth]{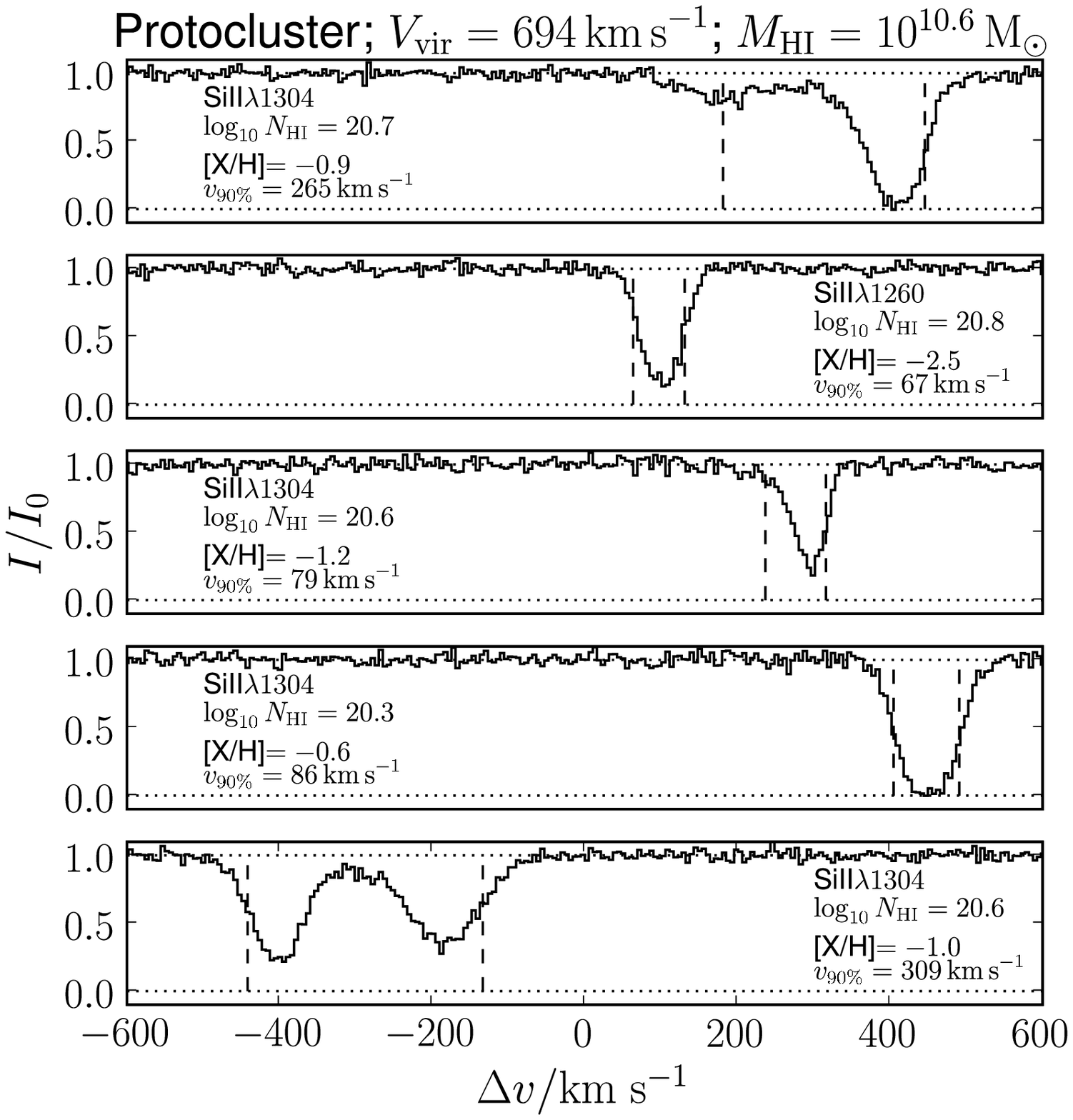}\includegraphics[width=0.48\textwidth]{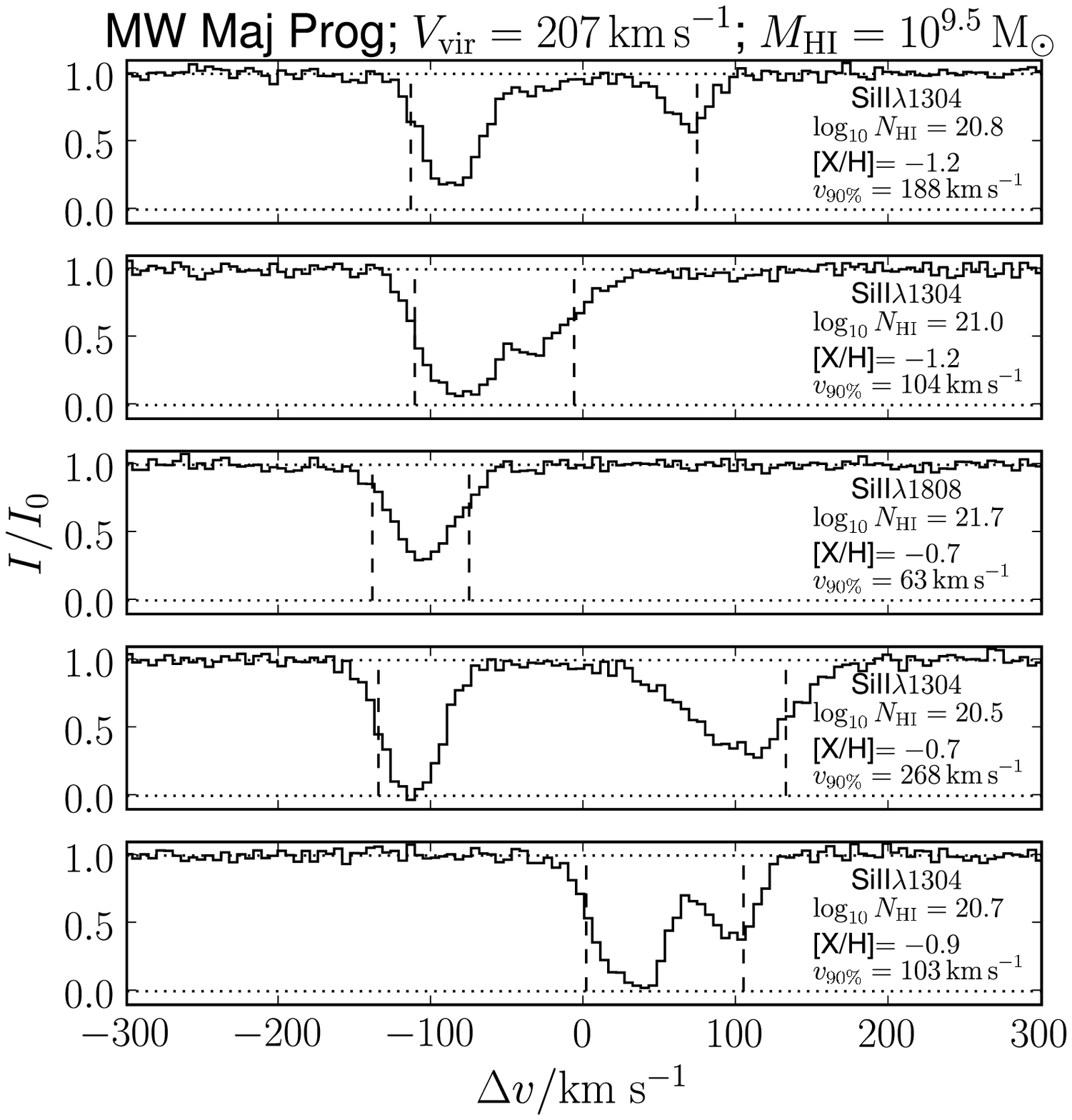}
\includegraphics[width=0.48\textwidth]{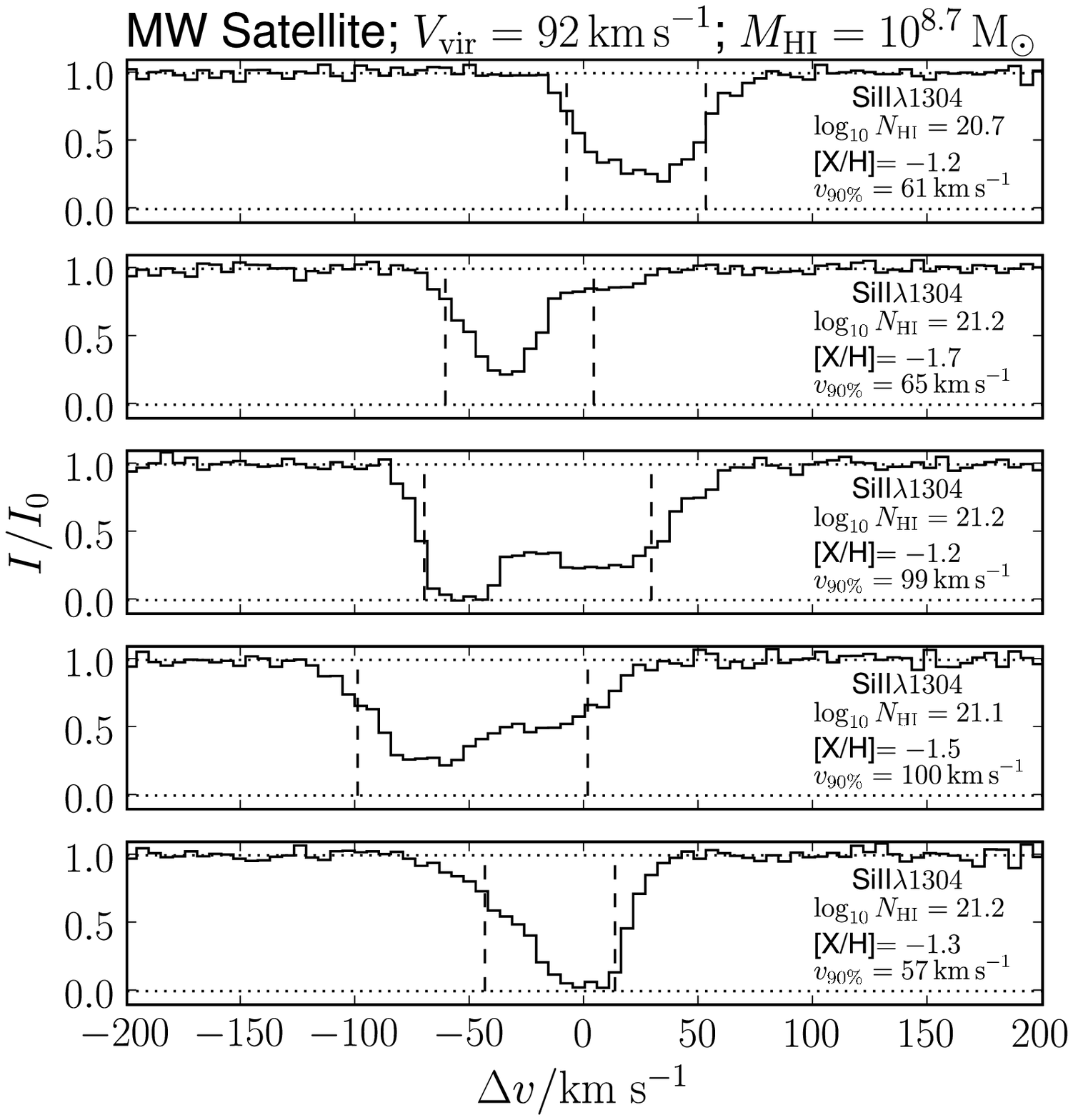}\includegraphics[width=0.48\textwidth]{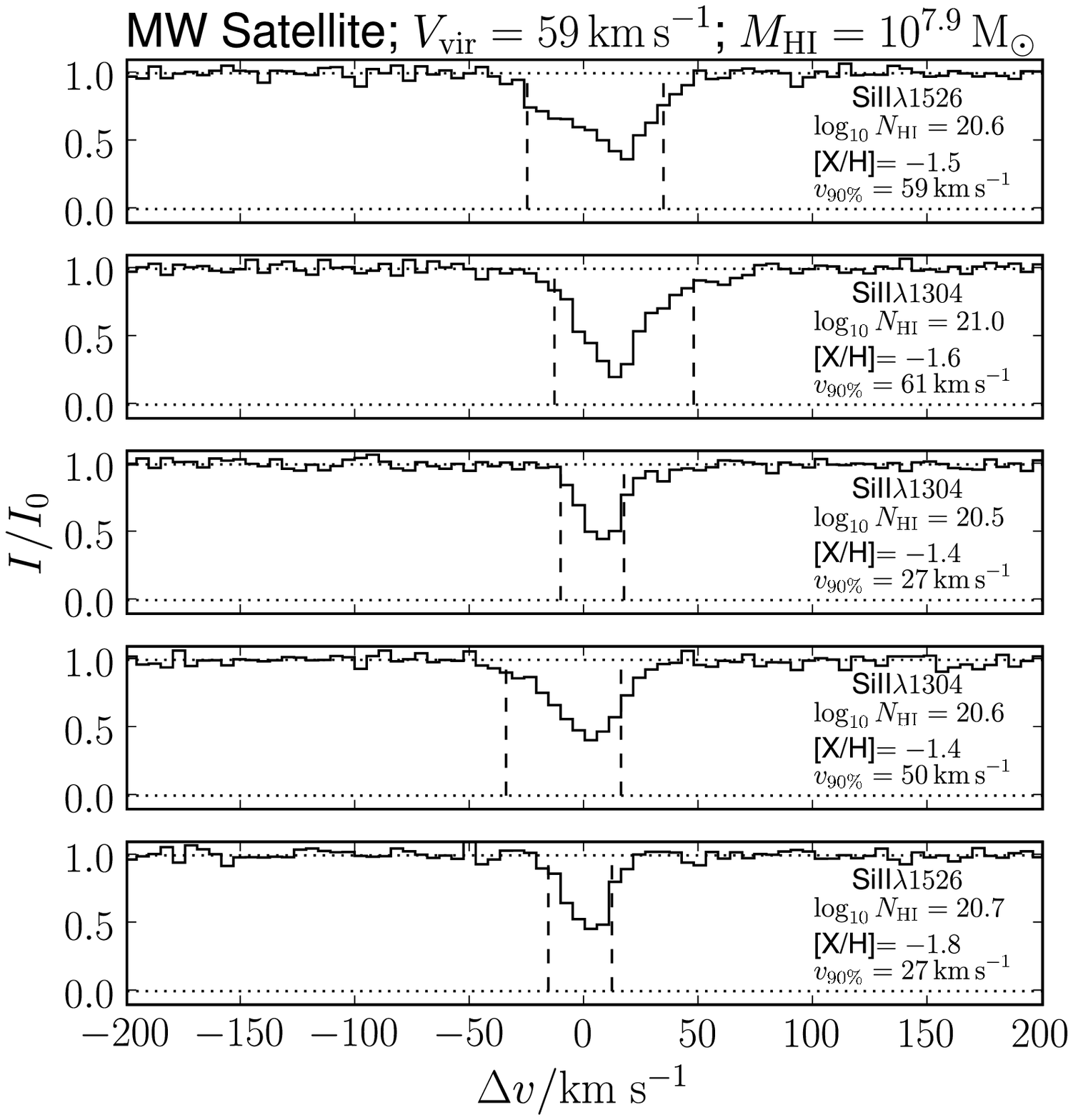}
\vspace{0.5cm}
\caption{Example low-ion (Si\,\textsc{ii}) absorption profiles from
  random sightlines through selected $z=3$ haloes: in reading order,
  these have virial velocities of $694$, $207$, $92$, $59 \kms$
  (making the first halo extremely rare in cosmological sightline
  samples, see Section \ref{sec:halo-properties}). The first is taken
  from our ``Cosmo'' box, the second is the ``MW'' major progenitor
  and the remaining two are smaller haloes from the ``MW'' simulation.
  Note that the velocity axes are scaled differently in the respective
  plots. The zero velocity offset corresponds to the motion of the
  centre of mass of the halo (unobservable in practice), illustrating
  a qualitative shift in behaviour from multiple clumps (higher virial
  velocities) to single clumps (lower virial velocities). The vertical
  dotted lines indicate the velocity offsets where the cumulative
  optical depth reaches $5\%$ and $95\%$ of its maximum value;
  $v_{90\%}$ is given by the difference in their position in velocity
  space. The pixel size is approximately $5 \kms$, although internally
  we use a higher $1 \kms$ resolution. For the purposes of this plot,
  the profiles are normalized to correspond to a definite transition,
  although this is not necessary for our computations.  This chosen
  transition, along with the sightline \hi~ column density and
  metallicity, is indicated in each panel.  Noise is added to simulate
  S:N $= 30:1$; again this is only for illustrative purposes and is
  not part of our pipeline. }\label{fig:lowion-profiles}
\end{figure*}

Line-of-sight properties can be calculated from particle-based
simulations by projecting quantities onto a grid. For instance,
projecting all gas particles in a simulation onto the $x$ -- $y$ plane
allows the column density along the $z$ direction to be estimated by
summing the mass in a grid square and dividing by its area. This is
the approach taken in previous SPH simulations of DLA
properties. However, in initial numerical experiments we found that
sightline properties in our simulations were not robust to changes in
the somewhat arbitrary grid resolution.

We have instead calculated all quantities using a true SPH
approximation; for details, see Appendix \ref{sec:sph-calculations}.
This results in a varying spatial resolution of sightlines which is
automatically consistent with the simulation data. The minimum
smoothing length allowed is $0.1$ times the softening (given in Table
\ref{tab:simulations}), meaning we can resolve spatial gradients over
as little as $\sim 20 \pc$ in our highest resolution simulation,
although a more typical effective resolution is $\sim 200 \pc$.

For our main results, sight-lines are projected through the simulation
in random orientations and at random sky-projected offsets from the
centre of a halo up to its virial radius. We verified that extending
this search area to twice the virial radius had no impact on our
results (this can also be seen directly from Figure
\ref{fig:projected-hi}). This confinement of our DLA cross sections is
discussed in Section \ref{sec:no-interg-dlas}.

For each sightline, we measure directly the column density in neutral
hydrogen. If this column density exceeds the DLA threshold
($N_{\mathrm{HI}}>10^{20.3} \cm^{-2}$), it is added to our catalogue.
If not, it is immediately discarded; but we keep track of the numbers
of all sightlines taken so that we may calculate
\begin{equation}
  \sigma_{\mathrm{DLA}} \equiv \sigma_{\mathrm{search}} \left( \frac{n_{\mathrm{DLA}}}{n_{\mathrm{total}}} \right)
\end{equation}
where $\sigma_{\mathrm{search}} = \pi r_{\mathrm{vir}}^2$ is the
search area, $n_{\mathrm{total}}$ is the total number of random
sightlines calculated and $n_{\mathrm{DLA}}$ is the number of such
sightlines which exceed the DLA threshold. In this way, we obtain
a representative DLA cross-section for each halo without assuming
any particular projection.

We also produce an absorption line profile for a low-ion transition
such as those of \siii. We assumed that the relative abundances of
heavy elements were solar and that \siii ~ was perfectly coupled to
\hi, so that for solar metallicity $M_{X}/M_{H}=0.0133$ and
$n(\textrm{Si\,{\sc ii}})/n(\textrm{H\,{\sc
    i}})=n(\textrm{Si})/n(\textrm{H})=3.47 \times 10^{-5}$
\citep{2003ApJ...591.1220L}. In other words, given the metallicity $Z$
of each gas particle, we take $n(\textrm{Si\,{\sc ii}}) = 3.47 \times
10^{-5} \, (Z/Z_{\mathrm{\odot}}) \, n(\textrm{H\,{\sc i}})$. Although
an approximation, we found that the effect of relaxing the assumption
of the \siii~ -- \hi~ coupling was minor; see Section
\ref{sec:where-is-siii}.

Example profiles are shown in Figure \ref{fig:lowion-profiles}, in
which we have chosen four haloes and displayed five random sightlines
from each. For the purposes of this plot, we choose one of
SiII$\lambda$1808, 1526, 1304 or 1260 according to which transition
has maximum optical depth closest to unity. The plots are centred such
that $\Delta v = 0$ corresponds to the motion of the centre of mass of
the parent dark matter halo. We also added, for display purposes,
gaussian noise such that the signal-to-noise ratio (S/N) is
$30/1$. The range contributing to $v_{90\%}$ (see Section
\ref{sec:veloc-width-distr}) is shown in Figure
\ref{fig:lowion-profiles} by vertical lines at either end.  Absorption
arising in haloes with virial velocities $v_{\mathrm{vir}}\gsim 150
\kms$ is often composed of multiple clumps whereas for smaller haloes
there tends to be one main, central \hi~ clump, moving with the centre
of mass of the halo. Visual inspection of the different systems
suggests that the former systems are less dynamically relaxed,
presumably because they have formed more recently and have longer
dynamical timescales; however, we did not verify this systematically.

\subsection{Cosmological Sampling}\label{sec:cosm-conv}

We aim to make statements about the agreement or otherwise of our
simulations with cosmological observations of DLAs and therefore need
to construct a representative global sample of absorbers.  Our
approach is related to that of \cite{1997ApJ...484...31G}, in that we
correct our limited sample by reweighting in accordance with the halo
mass function. However, we emphasize that in our case this is to make
allowance for our limited statistics and combine results from separate
boxes, whereas in \cite{1997ApJ...484...31G} this approach was used to
extrapolate behaviour into an unresolved low-mass regime.

Consider any measurable property of DLA absorbers, $p$. The aim of the
process discussed below is to construct the distribution function,
$\dd^2 N/ \dd X \dd p$. (We will adopt the custom of using the {\it
  absorption distance} $X$, where $\dd X/\dd z = H_0 (1+z)^2 / H(z)$:
populations with constant physical cross sections and comoving number
densities maintain constant line densities $\dd N/\dd X$ as they
passively evolve with redshift.) Given the monotonic invertible
relations $p(\mvir)$ and $\sdla(\mvir)$, one has

\begin{equation}
\frac{\dd^2 N}{\dd X \dd p} = f[\mvir(p)] \sdla[\mvir(p)] \frac{\dd l}{\dd X} \frac{\dd \mvir}{\dd p}\label{eq:std-convolve}
\end{equation}
where $\sigma_{\mathrm{DLA}}(M)$ is the cross section of a DLA of mass
$M$ and $f(M)$ is the halo mass function\footnote{We adopt the
  numerically calibrated version of the \cite{1999MNRAS.308..119S}
  halo mass function given by \cite{2006astro.ph..7150R}.}, so that
the physical number density of haloes with mass in the range $M \to
M+\delta M$ is $f(M)\delta M$, and $\dd l / \dd X= c / H_0 (1+z)^3$.
However, for an ensemble of haloes of the same mass, $\sdla$ will be
scattered -- although it may be possible to define a mean value
$\bar{\sigma}_{\mathrm{DLA}}$. Similarly, $p$ will in fact be
scattered around some suitably chosen average $\bar{p}(M)$ for a given
halo mass; furthermore even for a {\it single} halo most properties
$p$ will vary depending on the particular sightline taken through the
halo to the distant quasar\footnote{We regard both of these effects as
  stochastic scatter, although presumably a complete theory would
  account for the exact value of $p$ in terms of a sufficient number
  of parameters pertaining to the halo and sightline.}.

The easiest approach is to make the replacements $p \to
\bar{p}(\mvir)$, $\sigma \to \bar{\sigma}(\mvir)$ in
equation~(\ref{eq:std-convolve}). However, even if significant
deviation of $p$ from $\bar{p}$ is rare, one may be interested in
variations of $\dd^2N/\dd X \dd p$ over several orders of magnitude --
these rare fluctuations can therefore contribute significantly to the
``tail'' of the observed values.

Accordingly, the method by which we perform the reweighting is
non-parametric. Since at first this can seem a little obscure, we
offer two descriptions. Below, we have described the method in a
heuristic manner. In Appendix \ref{sec:conv-meth-discr} we have
outlined how this method can be derived by discretizing a well defined
integral, which allows for an exact interpretation of our results
should one be necessary.

First, we bin haloes (in logarithmic bins) by their virial mass. Within
each bin $i$, ranging over virial mass $M_i \to M_{i+1}$, one may
calculate a mean DLA cross-section $\bar{\sigma}_i$ and a total
physical number density $F_i$:

\begin{equation}
F_i = \int_{M_i}^{M_{i+1}} f(M) \dd M \textrm{\,.}\label{eq:F-binned}
\end{equation}

The contribution of DLA systems per unit physical
length from the bin $i$ is $F_i \bar{\sigma}_i$, and consequently one
may calculate:

\begin{equation}
\frac{\dd N}{\dd X} = \frac{\dd l}{\dd X} \sum_i F_i \bar{\sigma}_i \textrm{ .}\label{eq:line-den-convolve}
\end{equation} 

To investigate the distribution of properties observed we have two
approaches. The simplest is to extract a representative set by
discarding selected sightlines. This is ideal for comparing
correlations between sightline parameters (e.g.  velocity \&
metallicity, Section \ref{sec:metallicity}), where the observed data
sets consist of only $\mathcal{O}(100)$ separate sightlines. For each
halo $h$, we need to select a number of sightlines proportional to
$w_h \equiv \sigma_h F_{i(h)}/n_{i(h)}$ (cf equation (\ref{eq:wh}))
where $\sigma_h$ is the cross-section of the specific halo, $i(h)$
represents the mass bin to which halo $h$ belongs and $n_i$ is the
total number of simulated haloes in the mass bin $i$.  The actual
sightlines chosen from each halo are determined by a pseudo-random but
deterministic approach, for stability of results.

\begin{figure}
\includegraphics[width=0.48\textwidth]{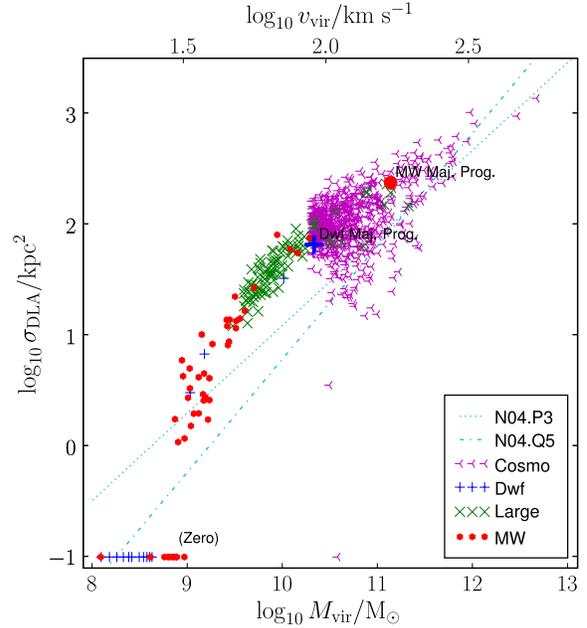}
\caption{The DLA cross-section of haloes which meet our resolution
  criteria in the Dwf (plus symbols), MW (dots), Large (cross symbols)
  and Cosmo (tripod symbols) boxes plotted against their virial
  mass. There is a (resolution-independent) sharp cut-off at
  $M_{\mathrm{vir}} \sim 10^9 \msol$ below which the cross-section for
  DLA absorption is negligible. Haloes with no DLA cross-section are
  shown artificially at $\log_{10} \sigma_{\mathrm{DLA}}/ \kpc^2 =
  -1$. The fit to the equivalent results for two models in
  \protect\cite{2004MNRAS.348..421N} is given by the dotted and
  dash-dotted lines (their models P3 and Q5 respectively). The major
  progenitors to the ``MW'' (Milky Way like) and ``Dwf'' (Dwarf type)
  $z=0$ galaxies are indicated.  }\label{fig:xsec-raw}
\end{figure}

This method is not ideal, however, because it throws away potentially
useful information. In particular, when constructing distribution
functions for a property $p$, we use an alternative method which uses
the cosmological halo mass function to weight, rather than select,
results.  The set of all sightlines is binned by the property $p$,
indexed by $j$. One then has

\begin{eqnarray}
\left. \frac{\dd^2 N_{\mathrm{DLA}}}{\dd X \dd p} \right|_{p_j} 
 \propto \nonumber
\end{eqnarray}
\begin{equation}
\hspace{7mm} \sum_{h} \frac{w_{h(k)}}{\Delta p}   \times 
\frac{\textrm{Num. DLA obs. through }h\textrm{ with }p\textrm{ in bin }j}{\textrm{Total DLA obs. through  }h}
\end{equation}
where the sum is over {\it all} sightlines from any halo in any box,
$h(k)$ denotes the halo associated with the sightline $k$ and $\Delta
p$ is the bin size.  The constants are not hard to calculate, but
obscure the technique and are presented in full in Appendix
\ref{sec:conv-meth-discr}, equation
(\ref{eq:discrete-p-final-binned}).

Of course, the above methods require that line-of-sight effects (such
as the coalignment of multiple DLAs along a single sightline) are not
a dominant effect. We verified that extending all our sightlines
through the entire boxes made no significant difference to any of our
results. This is because the DLA cross-sections per halo are small so
that even when strong clustering is taken into account, the
probability of a double-intersection along a line-of-sight is very
low. (Note also that velocity widths, Section
\ref{sec:veloc-width-distr}, are always determined from unsaturated
line profiles, so that trace metals in the IGM are too weak to change
the measured widths.)

One should also require that environmental variations of halo
properties (in effect systematic variations with parameters other than
halo mass) are unimportant. This requirement is harder to verify, but
we neither detect variations of DLA properties of differing regions in
our ``Cosmo'' box, nor any strong correlations with indirect measures
of environment such as the halo spin parameters. More systematic work
in the form of the \textsc{Gimic} project, resimulating carefully
chosen regions of the Millennium Simulation
\citep{2005Natur.435..629S}, also suggests that environmental
considerations are largely controlled by the local variation of the
finite volume halo mass function (Crain et al., in prep).

\section{Results}\label{sec:results}

\subsection{Masses of Haloes Hosting DLAs}\label{sec:halo-properties}
 
\begin{figure}
  \includegraphics[width=0.48\textwidth]{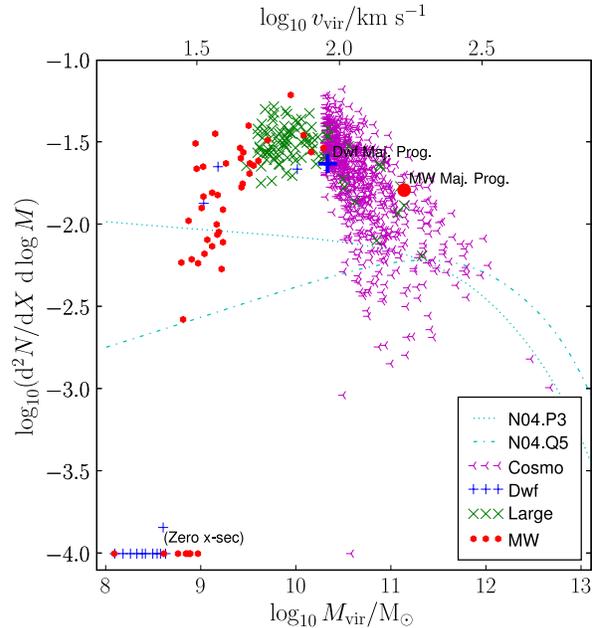} 
  \caption{The data of Figure \ref{fig:xsec-raw} multiplied by the
    halo mass function from \protect\cite{2006astro.ph..7150R} to give
    a total line density for each representative system. Haloes with
    zero cross-section are shown artificially at $\log_{10} \dd^2N/\dd
    X \dd \log_{10} M=-4$. }\label{fig:xsec-hmf}
\end{figure}

Taking all haloes of sufficient resolution (Section
\ref{sec:resolution}), we plot their DLA cross-section against virial
mass in Figure~\ref{fig:xsec-raw}. Our DLAs have cross-sections of
between $1$ and $1\, 000 \kpc^2$.  In agreement with rough physical
expectations, the cross-section increases as a function of mass.  In
their regions of overlap, the results agree between boxes: although
the Cosmo box has a vastly larger volume than our other boxes, and
hence exhibits more scatter by probing rarer haloes, we verified that
by restricting the number of haloes at a given mass the distributions
of haloes from differing simulations were in agreement in each mass
bin.  Two extreme outliers in the Cosmo box were individually
inspected and found to be systems in the middle of major mergers.

\begin{figure}
\includegraphics[width=0.48\textwidth]{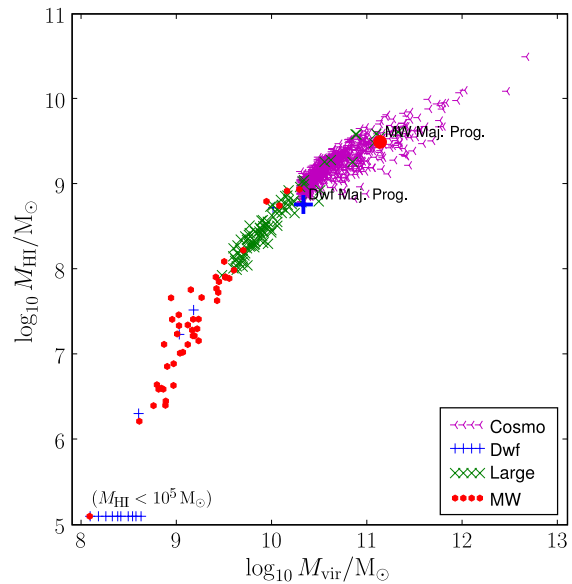}
\caption{The total mass of neutral hydrogen plotted against the virial
  mass for our haloes. The symbols are as described in the caption of
  Figure \ref{fig:xsec-raw}.}
\label{fig:hi-vs-mvir}
\end{figure}

A notable feature is that the cross-section drops off extremely
quickly for $M_{\mathrm{vir}}<10^9 \msol$
(i.e. $v_{\mathrm{vir}}(z=3)<30 \kms$). This behaviour is accompanied
by a sharp break in the mass of neutral hydrogen in such haloes (Figure
\ref{fig:hi-vs-mvir}), and we attributed it to the UV field which
prohibits cooling of gas in lower mass haloes \citep[see][ and Section
\ref{sec:no-interg-dlas}]{1986MNRAS.218P..25R,1992MNRAS.256P..43E,1996MNRAS.278L..49Q}.

Our cross-sections are overall somewhat larger than previous
simulation works have suggested
\cite[e.g.][]{2004MNRAS.348..421N,1997ApJ...486...42G} and furthermore
are not compatible with a single power-law link to the halo mass. The
results of two models from \cite{2004MNRAS.348..421N} are overplotted
on Figures \ref{fig:xsec-raw} and \ref{fig:xsec-hmf} for comparison
(runs ``P3'' and ``Q5'' here refer to weak and strong galactic winds
in the cited work). Our enhanced cross-sections for $M_{\mathrm{vir}}
\sim 10^{10} \msol$ are apparently a consequence of our particular
feedback implementation -- see Section \ref{sec:feedback}.

From our cross-sections, we can calculate directly the line density of
DLAs $l_{\mathrm{DLA}}\equiv\dd N / \dd X$ via equation
(\ref{eq:line-den-convolve}). For this, we find
$l_{\mathrm{DLA}}=0.070$, in good agreement with the observationally
determined value $l_{\mathrm{DLA}}(z=3)= 0.065 \pm 0.005$ from SDSS
DR5\footnote{\texttt{www.ucolick.org/\~{}xavier/SDSSDLA/DR5/}} using
the method described for SDSS DR3 in \cite{2005ApJ...635..123P}.  In
Figure \ref{fig:xsec-hmf}, we have shown how haloes of different masses
contribute to this total line density by plotting $\dd^2 N/ \dd X \dd
\log_{10} M = (\ln 10)\left(\dd l / \dd X\right) M f(M)
\sigma_{\mathrm{DLA}} $ for each halo; $l_{\mathrm{DLA}}$ is simply
the integral under the curve defined by the locus of these points. The
major contributors to the total line density are haloes of masses $10^9
\msol < M_{\mathrm{vir}} < 10^{11} \msol$. At lower and higher masses
the contribution is cut off by the rapidly decreasing cross-sections
or exponential roll-off in the halo mass function respectively.

As expected from our previous discussion, our results have a peak at
$\sim 10^{10} \msol$ which contrasts with the flatter results from
fiducial power-law cross-sections. Consequently, at first glance, it
appears that the area under our locus of points must be larger than
that under the N04 curves and hence a substantial disagreement in line
density is inevitable; however the cut-off for N04 is at rather lower
masses ($M_{\mathrm{vir}} \sim 10^8 \msol$), and this brings the total
line density in N04 considerably closer to the observed value.

\begin{figure}
\begin{center}
\vspace{0.5cm}
\includegraphics[width=0.43\textwidth]{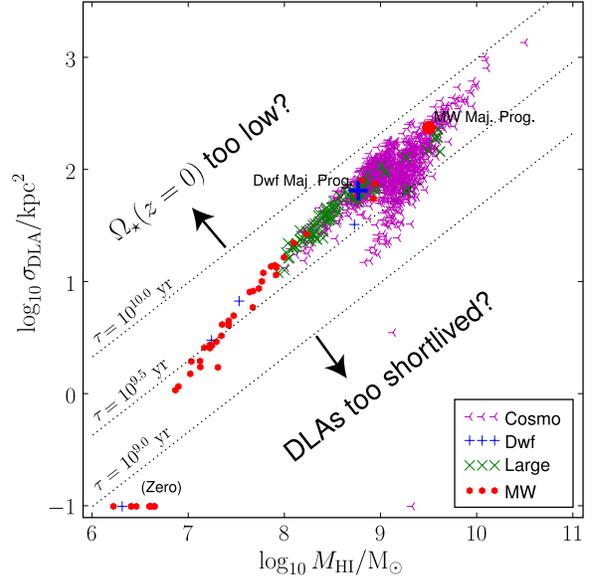}
\end{center}
\caption{ As Figure \ref{fig:xsec-raw}, except the cross-sections are
  now plotted against the total mass of neutral hydrogen in each
  halo. Haloes with no DLA cross-section are shown artificially at
  $\log_{10} \sigma_{\mathrm{DLA}}/ \kpc^2 = -1$. The dotted lines are
  of timescales for \hi~ depletion through star formation; from top to
  bottom $\tau = 10^{10.0}$, $10^{9.5}$ and $10^{9.0}$ yr. These
  illustrate constraints on our locus of points by considering both
  $\Omega_{\star}(z=0)$ and the lifetime of a typical DLA (see text
  for details).}
\label{fig:xsec-vs-hi}
\end{figure}

Plotting the total \hi~mass against the virial mass (Figure
\ref{fig:hi-vs-mvir}) and DLA size against the total \hi~mass of a
halo (Figure \ref{fig:xsec-vs-hi}) gives an alternative view of our
cross-sections. A striking feature of the latter plot is a
bifurcation, particularly notable in the ``Cosmo'' box but also traced
by the ``Large'' box, in which haloes of a fixed \hi~mass
$M_{\mathrm{HI}} \sim 10^{9} \msol$ can have different
cross-sections. The primary physical distinction between the upper and
lower branches is in halo mass: the former trace \hi-rich haloes with
$M_{\mathrm{vir}}<10^{10.5} \msol$ and the latter trace a population
of \hi-poor haloes with $M_{\mathrm{vir}}>10^{10.5} \msol$. This is
reminiscent of recent claims of bimodality in observed DLAs
\citep{2008arXiv0802.3914W}. However we limit ourselves, for the
moment, to more general considerations, noting that the high mass, low
$\sigma$ branch DLAs are extremely rare in our simulations (making up
less than 2\% of the total cross-section). Our method for generating
cosmological samples (Section \ref{sec:cosm-conv}) will propagate
through any such bimodalities into our final results without
difficulty, assuming the Cosmo box has a representative selection of
haloes. See Section \ref{sec:bimodality} for a further discussion of
bimodality.

A guide $\sigma_{\mathrm{DLA}}$ -- $M_{\mathrm{HI}}$ relationship can
be estimated by fixing a particular star formation rate. As explained
in Section \ref{sec:simulations}, in our simulations the instantaneous
star formation rate is given by the Schmidt law. From this may be estimated
a timescale for conversion of neutral gas into stars,
\begin{eqnarray}
\tau = \frac{1}{c_{\star} \sqrt{G \rho_{\mathrm{gas}}}} \textrm{ .}
\end{eqnarray}
With the assumption that
$\rho_{\mathrm{gas}} \simeq M_{\mathrm{HI}}/(0.74 ~
\sigma_{\mathrm{DLA}}^{3/2})$, one has the approximate relation
\begin{equation}
\sigma_{\mathrm{DLA}} \sim 10 \kpc^2 \left(\frac{\tau}{10^9 \yr}\right)^{4/3} \left(\frac{M_{\mathrm{HI}}}{10^9 \msol}\right)^{2/3}\left(\frac{c_{\star}}{0.05}\right)^{4/3} \textrm{.}
\end{equation}

Our locus lies roughly along the line $\tau = 10^{9.5}$ years; this is
not unexpected, especially if our simulations are to match
observations that $\Omega_{\star}(z=0) \sim
\Omega_{\mathrm{DLA}}(z=3)$.\footnote{In other words, the total mass
  of stars at $z=0$ in a given comoving volume is roughly equal to the
  total mass of neutral hydrogen in that same volume at $z=3$; see
  equation (\ref{eq:omega-dla}) and the ensuing discussion.} This
suggests that a large fraction of the DLA cross-section should be
converted to stars by $z=0$, giving an upper limit of $\tau \lsim
\mathcal{O}(10^{10} \yr)$ to the star formation timescale $\tau$
(unless star formation proceeds in rapid discrete bursts, which is not
the case in our simulations).  Assuming DLAs are not short-lived
objects, or achieved by very fine balancing of rapid gas cooling and
star formation, one would also expect $ \tau \gsim
\mathcal{O}(1/H(z=3)) \simeq \mathcal{O}(10^{9.5} \yr)$. The
constraint $\tau > 10^9 \yr$ is obeyed, suggesting that this stable
model is reasonable.
\subsection{Column Densities, Velocity Widths and Metallicities}

\subsubsection{Column Density Distribution}\label{sec:column-dens-distr}

\begin{figure}
\includegraphics[width=0.48\textwidth]{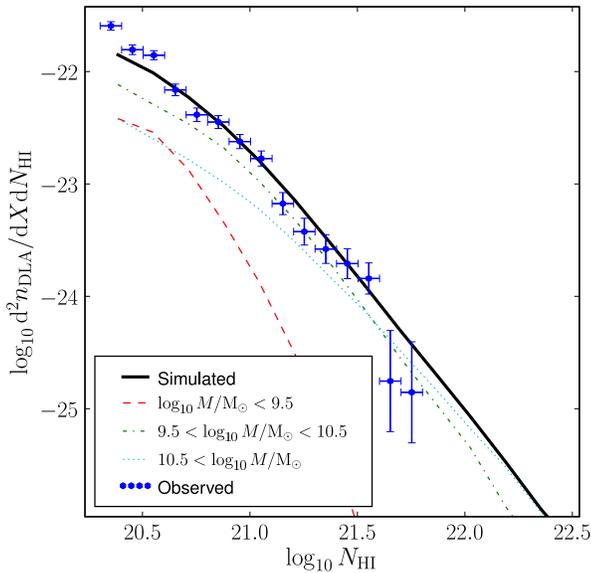}
\caption{The simulations' DLA column density distribution (solid line)
  compared to the observed values from SDSS DR5 (points with error
  bars, based on \protect\citeauthor{2005ApJ...635..123P} 2005; see
  main text for explanation). The dashed, dashed-dotted and dotted
  lines show the contribution from $M_{\mathrm{vir}}<10^{9.5} \msol$,
  $10^{9.5} \msol<M_{\mathrm{vir}}<10^{11.0} \msol$ and
  $M_{\mathrm{vir}}>10^{11.0} \msol$ haloes respectively (these are not
  directly observable distributions, but give guidance as to how our
  cross-section is composed). }\label{fig:colden}
\end{figure}

One of the best constrained quantities, observationally, is the
neutral hydrogen column density distribution $f(N_{\himath},X)$. This
is defined such that $f(N_{\himath},X) \dd N_{\himath} \dd X$ gives
the number of absorbers with column densities in the range
$N_{\himath} \to N_{\himath} + \dd N_{\himath}$ and absorption
distance $X \to X + \dd X$. Applying the reweighting method described
in Section \ref{sec:cosm-conv} to our sample yields an estimate for
the cosmological column density distribution, shown by the solid line
in Figure \ref{fig:colden}. This can be compared directly to the
observed distribution given by the points with error bars, which are
derived from SDSS DR5 (see previous section for an explanation). The
matching of the normalization and approximate slope of the observed
column density distribution can be seen as a genuine success of the
simulations: we emphasize that no fine tuning has been applied to
achieve this result.  Furthermore, our results appear to have
converged at the resolution of the simulations used (see Section
\ref{sec:resolution}).

From the column density distribution, one may express the total
neutral gas mass in DLAs in terms of the fiducial definition
\begin{equation}
\Omega_{\mathrm{DLA}}(z) = \frac{m_pH_0}{c f_{\mathrm{HI}}\rho_{c,0}} \int_{10^{20.3}\,\cm^{-2}}^{N_{\mathrm{max}}} f(N_{\himath},X) N_{\himath} \dd N_{\himath} \label{eq:omega-dla}
\end{equation}
where $m_p$ is the proton mass, $\rho_{c,0}$ is the critical density
today, $(1 - f_{\mathrm{HI}}) \simeq 0.24$ gives the fraction of the gas
in elements heavier than hydrogen and $N_{\mathrm{max}}$ is an upper
limit for the integration, which is discussed in the next
paragraph. $\Omega_{\mathrm{DLA}}(z)$ gives the fraction of the
redshift zero critical density provided by the comoving density of DLA
associated gas measured at redshift $z$. (This is different from the
more natural definition of time-dependent $\Omega$s which express a
density at any given redshift in terms of the critical density at that
redshift. Only in the Einstein-deSitter universe will these
definitions coincide.)

Although the calculation should take $N_{\mathrm{max}} = \infty$, this
is not possible for the observational sample owing to the rapidly
decreasing number of systems at the high column density
limit. \cite{2005ApJ...635..123P} discussed how different assumptions
for the functional form of the column density distribution can lead to
different values of $\Omega_{\mathrm{DLA}}$. The discrepancies are
small for the two best functional fits to the observational data (a
double power law or a Schechter function with exponential roll-off at
high column densities). However, these extrapolations are actually
only constrained by a few points at high column densities; a more
robust approach -- albeit less physically transparent -- is to
calculate $\Omega_{\mathrm{DLA}}$ directly from summing the total
neutral hydrogen in the observed sample of DLAs, which for the SDSS
DR5 sample is roughly equivalent to using the upper limit
$N_{\mathrm{max}}=10^{21.75} \cm^{-2}$.

Using this limit, we obtain $\Omega_{\mathrm{DLA,sim}} = 1.0 \times
10^{-3}$, which can be compared with the result from SDSS DR5 in the
combined bin $2.8<z<3.5$, $\Omega_{\mathrm{DLA,obs}} = (0.84 \pm 0.06)
\times 10^{-3}$. As expected from Figure \ref{fig:colden}, the results
are in fair agreement; the slight mismatch is driven by the
overestimation of our high column density points ($10^{21.5} <
N_{\mathrm{HI}}/\cm^{-2} < 10^{21.75}$).

Our weighting approach predicts the form of the distribution for much
rarer, higher column density systems than the sample limited
observations allow.  Significant contributions to
$\Omega_{\mathrm{DLA}}$ are made by these rare systems unless $\gamma
\equiv \dd \ln f(N,X) / \dd \ln N \ll -2$.  In fact, directly
measuring the slope $\gamma$ for our simulations shows that it slowly
decreases from $\gamma \simeq -1.0$ for $N_{\mathrm{HI}} = 10^{20.3}
\cm^{-2}$ to a constant value of $\gamma \simeq -2.5$ for
$N_{\mathrm{HI}} \ge 10^{21.5} \cm^{-2}$. Thus a correction to
$\Omega_{\mathrm{DLA,sim}}$ is expected if we allow $N_{\mathrm{max}}$
to extend to arbitrarily high values. Performing the calculation with
$N_{\mathrm{max}}= \infty$ gives $\Omega_{\mathrm{DLA,sim}} = 1.4
\times 10^{-3}$. This value is not directly comparable to
observational estimates, but shows that an observer living in our
simulations would underestimate $\Omega_{\mathrm{DLA}}$ by about
$30\%$ due to missing contributions from the rare high column density
systems. A further discussion of this issue is given in Section
\ref{sec:column-dens-disc}.

\subsubsection{Velocity Width Distribution}\label{sec:veloc-width-distr}
\begin{figure}
\includegraphics[width=0.48\textwidth]{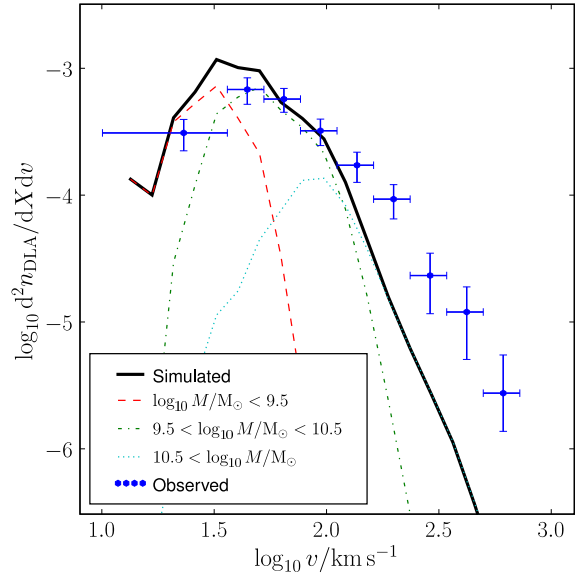}
\caption{The simulations' DLA velocity distribution (solid line)
  compared to the observed values based on the sample described by
  \protect\cite{2003ApJ...595L...9P} (see text for details; shown by
  points with error bars). The dashed, dashed-dotted and dotted lines
  are as described in the caption of Figure
  \ref{fig:colden}.}\label{fig:veldistrib-std}
\end{figure}

We have already discussed some qualitative features of the low-ion
velocity profiles generated in our simulations (Section
\ref{sec:sightline-projection}, Figure~\ref{fig:lowion-profiles}).
These are important because they provide a direct observational
measure of the kinematics of the DLAs, and therefore have the ability
to substantially constrain the nature of the host haloes.  We now turn
to the comparison of our characteristic velocities with the observed
quantitative distribution.

We assign a velocity width to each generated profile using the
fiducial $v_{90\%}$ technique \citep{1997ApJ...487...73P}.  This
inspects the ``integrated optical depth'' $T(\lambda) =
\int_0^{\lambda} \dd \lambda ' \tau(\lambda')$ and assigns the
velocity width $ v_{90\%} = c (\lambda_b-\lambda_a)/\lambda_0$ where
$T(\lambda_b) = 0.95 T(\infty)$ and $T(\lambda_a) = 0.05 T(\infty)$.
The result is a representative velocity width for the sightline,
produced without any of the difficulties associated with fitting
multiple Voigt profiles.

To a good approximation, the only dependence on the particular low-ion
transition chosen is in the overall normalization of the optical
depths from the relative abundances and oscillator strengths. The
$v_{90\%}$ measure of the velocity width is invariant under such
rescalings, so that only for display purposes (i.e. Figure
\ref{fig:lowion-profiles}) do we need to choose a particular ion and
transition. Observers require to choose unsaturated lines because,
while our simulations calculate $\tau$ directly, spectra only
determine $e^{-\tau}$ which cannot be inverted (in the presence of
noise) for $\tau \gg 1$.

Using our weighting technique, we calculate the cosmological
distribution of velocity widths, $\dd^2 N / \dd X \dd v$. This is
plotted in Figure \ref{fig:veldistrib-std} (solid line), along with
observational constraints (points with error bars) calculated from
data provided by Jason Prochaska (private communication, 2008) based
on the compilation of high resolution ESI, HIRES and
UVES\footnote{Echellette Spectrograph and Imager on the Keck
  Telescope; High Resolution Echelle Spectrometer on the Keck
  Telescope; Ultraviolet and Visual Echelle Spectrograph on the Very
  Large Telescope} spectra described in \cite{2003ApJ...595L...9P}.
The dataset consists of 113 observed DLAs with
$4.5>z_{\mathrm{DLA}}>1.6$ (we directly verified that the velocity
width redshift evolution over this range is negligible, so that we use
the full range of observed systems to bolster our statistics). We
normalize the line density of DLAs in the observational sample to
match the observed $l_{\mathrm{DLA}}(z=3)=0.065$ (Section
\ref{sec:column-dens-distr}).

\begin{figure}
\vspace{0.30cm}
\begin{center}
\includegraphics[width=0.48\textwidth]{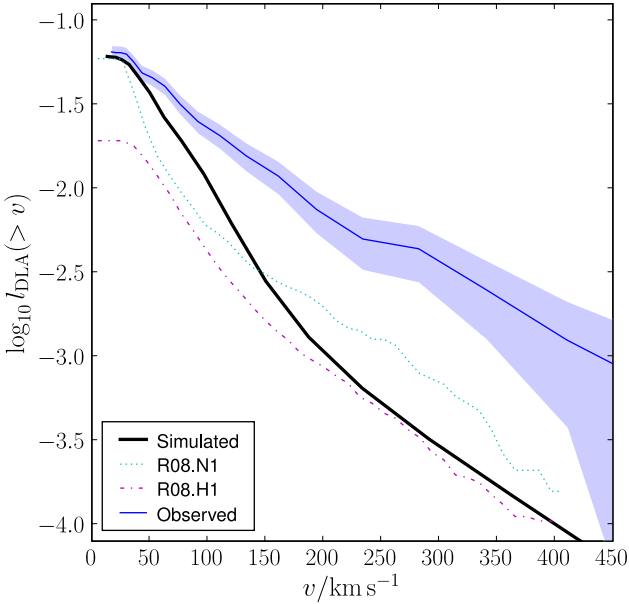}
\end{center}
\vspace{-0.5cm}
\caption{The data of Figure \ref{fig:veldistrib-std} (thick solid
  line) plotted cumulatively for comparison with similar plots in the
  literature. The dotted and dash-dotted lines show the
  \protect\cite{2007arXiv0710.4137R} models N1 and H1 respectively
  (see text for details). The shaded band gives the approximate
  Poisson errors on the observational data, but we caution these are
  in fact strongly correlated.}\label{fig:veldistrib-cum}
\end{figure}

Overall, the simulations reproduce the approximate pattern of observed
velocity widths, with a peak in the distribution at $v \sim 50 \kms$;
the agreement is fair for velocity widths $ v < 100 \kms$. However, it
is now a familiar feature of $\Lambda$CDM simulations that they
produce too few high velocity width systems (see Introduction) and our
results are no exception.  We provide a discussion of this discrepancy
in Section \ref{sec:velocity-profiles}.

Recent velocity width results by \cite{2007arXiv0710.4137R} and
\cite{Razoumov:2005bh} have been presented by plotting a cumulative
line density, $l_{\mathrm{DLA}}(>v)$.  A direct comparison between our
results and those from such work can be drawn from Figure
\ref{fig:veldistrib-cum}, in which we similarly replot our velocity
width distribution cumulatively (thick solid line). As well as the
observational data (thin solid line) we have overplotted results from
two simulations described by \citeauthor{2007arXiv0710.4137R} (2007;
R08). These two models, R08.N1 and R08.H1, differ in their size and
hence resolution; the former covers a volume of approximately $(5.7\,
\Mpc)^3$ (comoving), resolving grid elements of minimum side length
$90\, \pc$ (physical) while the corresponding values for the latter
are $(46 \, \Mpc)^3$ and $0.7 \kpc$. Comparing our simulations with
those of R08, the extent of the disagreement between simulated and
observed results is similar (discounting the lack of low velocity
systems in R08.H1, which arises from the coarser resolution), although
our own results are actually somewhat closer for low velocity widths
($v < 100 \kms$). This is possibly linked to our somewhat higher than
normal cross-sections for haloes of virial velocities
$v_{\mathrm{vir}} \sim 100 \kms$, but because much of R08's DLA
cross-section lies outside haloes it is hard to provide a concrete
explanation for such disparities (see Section
\ref{sec:no-interg-dlas}).

We caution that any cumulative measure has strongly correlated errors
so that the plot can present a somewhat distorted picture of the
discrepancy (one is really interested in its gradient in linear
space). In our plot we have shown the Poisson errors on the observed
data as a shaded band, but these only represent the diagonal part of
the covariance.

\subsubsection{Metallicity}\label{sec:metallicity}

\begin{figure}
\includegraphics[width=0.48\textwidth]{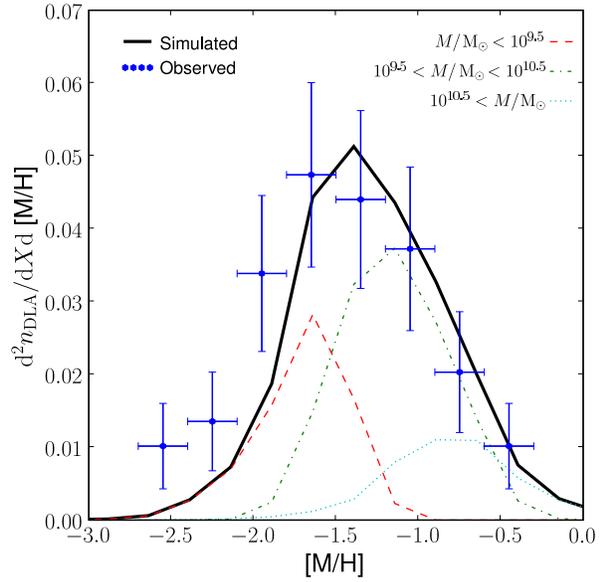}
\caption{The simulated distribution of metallicities (thick solid
  line). The points with error bars show the observed metallicities
  from the $2<z<4$ sample of DLAs based on
  \protect\cite{2007ApJS..171...29P}, normalized to the observed
  $l_{\mathrm{DLA}}$. The dashed, dashed-dotted and dotted lines
  are as described in the caption of Figure~\ref{fig:colden}.}\label{fig:metals}
\end{figure}

The metallicity, i.e. the ratio by mass of elements heavier than
helium to hydrogen, is an important diagnostic of observed DLA
sightlines. Since metals are deposited in the interstellar medium
(ISM) through supernova explosions, the metallicity of a region is
determined by the interplay between the integrated star formation rate
and bulk motions of gas including galactic inflows and outflows. In
general, one observes a positive correlation between the mass of a
galaxy and its metallicity both at $z=0$
\citep[e.g.][]{2004ApJ...613..898T,2006ApJ...647..970L} and at higher
redshifts \cite[e.g.][]{2005ApJ...635..260S,2006ApJ...644..813E}. This
trend is also observed in our simulations \citep[][ henceforth
B07]{2007ApJ...655L..17B}. Debate over a precise account of the origin
of the relation continues, but the analysis of B07 shows that, for our
simulations, the predominant effect is reduced star formation in low
mass haloes, with the dynamics of outflows providing an important
correction to the simple closed box model.

There are many uncertainties in simulating metallicities, which
reflect not only the mass formed in stars but also the dynamics of the
gas, its accumulation from the IGM and mixing within the halo. B07
showed that, for our simulations, high resolution is required
($N_{\mathrm{DM}}>3500$) to attain convergence for the star formation
history (and consequently the metallicities)\footnote{This requirement
  is weaker than the analagous result in recent work by
  \cite{2007ApJNaab} who require $10^5$ particles for convergence;
  this probably relates to the neglect of feedback effects in
  \cite{2007ApJNaab} -- feedback in our model tends to prevent
  collapse to very high densities, and thus imposes an effective star
  formation scale independent of the resolution; see also Section
  \ref{sec:feedback}.} .  We verified (using an independent analysis
code) that our own results suffered similarly, but that our other
results converge at lower resolutions (see Section
\ref{sec:resolution}). Whenever we deal with results concerning
metallicities, we therefore have to discard a number of haloes (those
with $N_{\mathrm{DM}}<3500$) which are included elsewhere. The effect
of this is to reduce the number of haloes in each mass bin, and
therefore exacerbate worries that we do not have a fully
representative set of haloes for all mass scales. Nonetheless, we do
not find any evidence that this is causing systematic effects, as we
verified that our other distributions are not significantly changed by
this restriction.

Each of our sightlines is assigned a metallicity as follows. The
simulations keep track of two sets of elements, the $\alpha$-capture
and Fe-peak elements. The enrichment process follows the model of
\cite{1996A&A...315..105R}, adopting yields for Type Ia and Type II
supernovae from \cite{1986A&A...158...17T} and
\cite{1993PhR...227...65W} respectively. For our purposes the
differences between the two metal groups are minor: observations show
corrections are typically $<0.3\,\mathrm{dex}$
\cite[e.g.][]{2002A&A...385..802L}, and exploratory work suggested our
simulations were similarly insensitive.  We choose to ignore these
differences for the sake of simplicity and use the total mass density
in metals. For quantitative results, we assume a solar metallicity
fraction of $0.0133$ by mass \citep{2003ApJ...591.1220L}.

A cosmological distribution of DLA metallicities is generated using
our fiducial technique (Section \ref{sec:cosm-conv}).  The result is
shown by the thick solid line in Figure \ref{fig:metals} and closely
matches the observational constraints, displayed as points with error
bars. These are derived from the \cite{2003ApJ...595L...9P} sample
described in Section \ref{sec:veloc-width-distr}, restricted to
$2<z<4$. We use metallicities based on $\alpha$-capture elements
(primarily Si), updated using data from \cite{2007ApJS..171...29P}.

Our simulated results are in good agreement with the observed
distribution. They exhibit a roughly gaussian distribution (as a
function of log metallicity); the best fit parameters give a median
metallicity of $[M/H]_{\mathrm{med,sim}} = -1.3$ with a standard
deviation $\sigma=0.45$. A similar fit to the observed data gives
$\mathrm{[M/H]}_{\mathrm{med,obs}} = -1.4$ and $\sigma=0.54$. Given
the uncertainties in abundances and yields, the differences are
extremely minor. This success is unusual: previous simulations
\cite[e.g.][]{2003ApJ...598..741C,2004MNRAS.348..435N} tended to
substantially overproduce metals in DLA systems, and were only been
able to approach the observed result if metal rich DLAs could be
hidden using substantial dust biasing.  However, such a scheme is
unsupported by observational evidence, and our simulations now suggest
it is unnecessary; for a further discussion see
Section~\ref{sec:column-dens-disc}.

We also investigated the relationship between the mean metallicity of
the gas within the virial radius of a halo and the spread of DLA
metallicities derived from SPH sightlines through that halo. We found
that the sightlines on average displayed slightly higher metallicities
than the halo mean (by $\lsim 0.5\, \mathrm{dex}$) with a spread of
$\sim 1\, \mathrm{dex}$. We interpret the offset by noting that DLA
sightlines sample the high density gas in which star formation is
occurring -- even if the rate is low (Section
\ref{sec:star-formation-rates}), presumably the local star formation
preferentially enriches these. Some haloes showed a detectable
radius-metallicity gradient, but in general this was shallow compared
with the overall scatter.

\subsubsection{Correlation between Metallicity and Velocity Widths}\label{sec:corr-betw-metall}

\begin{figure}
\includegraphics[width=0.48\textwidth]{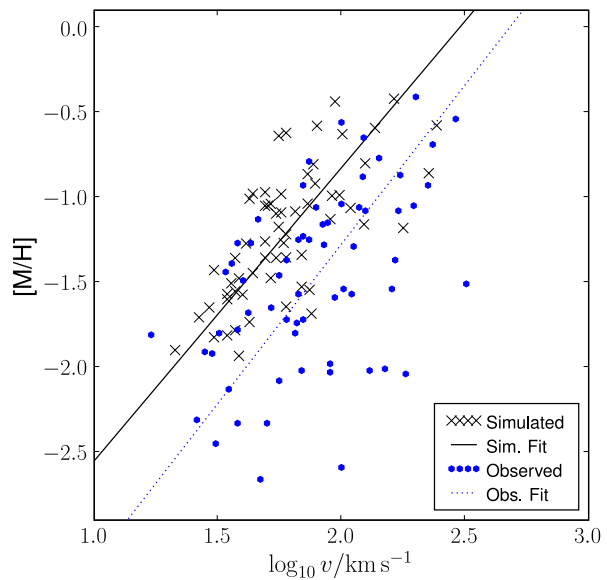}
\caption{The relationship between metallicity and low-ion velocity
  width of individual sightlines in a cosmologically weighted sample
  from our simulation (crosses, with linear least square bisector
  given by the solid line) and from the observational dataset
  described in the text (dots, with linear least square bisector given
  by the dotted line). The qualitative agreement, given the known
  deficiencies in the simulations and the lack of fine tuning, is
  remarkable. In the simulations, the origin of the correlation is an
  overall mass-metallicity relation; see text for further
  details. }\label{fig:velmetals}
\end{figure}

Observationally, there is known to be a correlation between the
low-ion velocity width and the metallicity of a DLA
sightline. \cite{2006A&A...457...71L} presented a set of observations
showing a positive correlation between these parameters at the $6
\sigma$ level and suggested that the relation could reflect a
mass-metallicity correlation analogous to that seen in galaxy surveys;
the result was confirmed, using a separate sample, by
\cite{2008ApJ...672...59P}.

Our matching of the metallicity relation and near-matching of the
velocity width distribution gives us confidence to attempt to probe
this relationship in our simulations. We employ the Monte-Carlo sample
generator version of our halo mass function correction code (Section
\ref{sec:cosm-conv}) to produce a sample of $64$ coupled velocity and
metallicity measurements, matching the size of the observational
comparison sample described in Section \ref{sec:veloc-width-distr}
restricted to $2<z<4$ as in Section \ref{sec:metallicity}.

The simulation results are shown as crosses in Figure
\ref{fig:velmetals}, with the linear least-square bisector
fit\footnote{The linear least-square bisector method estimates the
  slope of a relationship between $X$ and $Y$ as the geometric mean of
  the slopes obtained by regression of $X(Y)$ and $Y(X)$; see
  \protect\cite{1990ApJ...364..104I}.} given by the solid line.  The
observational sample is shown by dots (the errors on each observation
are small compared to the intrinsic scatter), with the linear
least-square bisector fit shown as a dotted line. These fitted
relationships are parametrized as $\log_{10} \Delta v_{\mathrm{sim}} =
2.5 + 0.58 \,[\mathrm{M/H}]$ and $\log_{10} \Delta v_{\mathrm{obs}} =
2.7 + 0.53 \, [\mathrm{M/H}]$ respectively.

Although the normalization of the relationship in our simulation is
somewhat different from the observational sample, we emphasize that
the slope and overall trend, as well as the mean metallicity of our
sightlines, are correct and that qualitatively the results are in
agreement. Given that the metallicity distribution (Figure
\ref{fig:metals}) is in close agreement with that observed, we suggest
that the quantitative disagreement arises from the previously noted
underestimate of the velocity widths in our simulation (i.e. one
should interpret the discrepancy in the relationships in Figure
\ref{fig:velmetals} as a horizontal, not vertical, displacement). It
is also apparent visually that the observational results show a larger
scatter about the mean relationship than does our computed sample. We
tentatively suggest that this points towards an ultimate resolution of
the velocity width issue which involves ``boosting'' the width of
certain sightlines, perhaps by local effects such as outflows, while
keeping the contribution by halo mass roughly as outlined in Section
\ref{sec:halo-properties} (although this is not the only conceivable
interpretation: see Section \ref{sec:velocity-profiles}).

The origin of our relation between velocity widths and metallicity is
the underlying mass-metallicity relation, as suggested by
\cite{2006A&A...457...71L}. We verified this directly, but it can also
be seen from Figures \ref{fig:veldistrib-std} and \ref{fig:metals}:
the dashed, dash-dotted and dotted lines in each case show the
contribution from haloes with $M_{\mathrm{vir}}<10^{9.5} \msol$,
$10^{9.5} \msol<M_{\mathrm{vir}}<10^{10.5} \msol$ and $10^{10.5} \msol
< \mvir$ respectively. Both the sightline velocities and the
metallicities can be seen to be a strong function of halo mass, and it
is this fact that leads to the final relationship.

\subsubsection{Other correlations are weak}

The correlation between column density and metallicity or velocity is
weak in our simulations, in agreement with observations: contrast the
strong dependencies of the velocity widths and metallicities on the
underlying halo mass with the weak dependence of the column density
(Figure \ref{fig:colden}).  Given the historical confusion over the
evidence for correlation between the column density and metallicity,
we have investigated this aspect of our observational and simulated
samples in a separate note (Pontzen \& Pettini, in preparation).

\subsection{Star formation rates}\label{sec:star-formation-rates}

\begin{figure}
\includegraphics[width=0.48\textwidth]{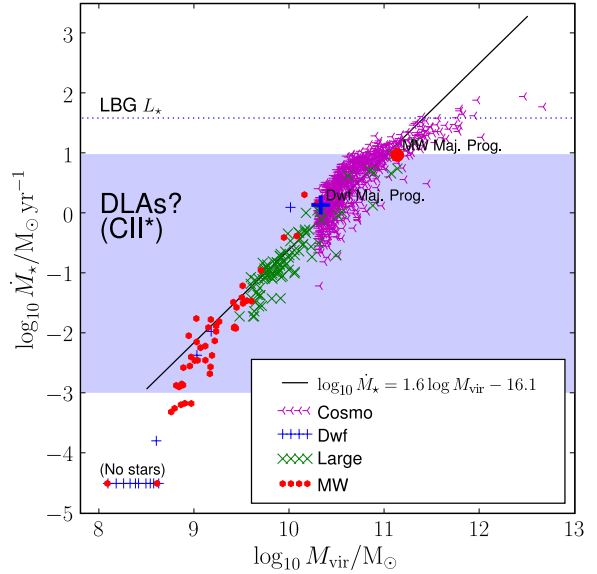}
\caption{The star formation rates in our haloes, plotted against their
  virial mass. Also shown is the observationally determined
  characteristic LBG star formation rate from
  \protect\cite{2007arXiv0706.4091R}, and a very approximate range of
  star formation rates for DLAs derived using the \cii cooling rate
  technique \protect\citep{2003ApJ...593..215W} -- for more details
  and a list of caveats, see main text. An empirical power law fit is
  shown. Haloes with no star particles are shown
  artificially at $\log_{10} \dot{M}_{\star} = -4.5$.}\label{fig:sfr}
\end{figure}

\begin{figure}
\includegraphics[width=0.48\textwidth]{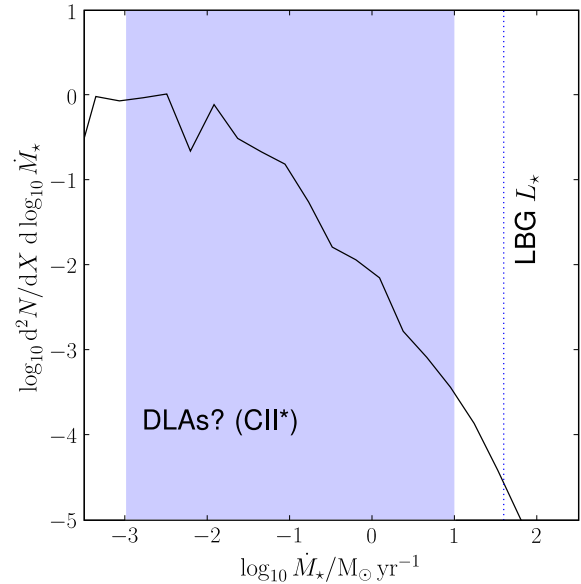}
\caption{Using our non-parametric weighting technique, the
  distribution of star formation rates in DLAs within our simulation
  is shown, with the same observed ranges indicated as in Figure
  \ref{fig:sfr}. There is qualitative agreement between our estimates
  for DLA star formation rates and the observational
  constraints.}\label{fig:dla-weighted-sfr}
\end{figure} 

In the Introduction, we noted that star formation rates (SFRs) in DLAs
are typically low ($\lsim 1 \msol \yr^{-1}$) and yet these systems
contain most of the neutral hydrogen, a necessary intermediary in the
star formation process. We also see this behaviour in our simulations,
and now describe how it comes about.

We associate a star formation rate with each halo by inspecting, at
$z=3$, the mass of star particles formed within the preceding $10^8$
years. For very low star formation rates
($<M_{\mathrm{p,gas}}/10^8\,\mathrm{yr}^{-1} \lsim 10^{-3} \mathrm{M}_{\odot}
\yr^{-1}$), there may have been no star particles formed in such a
period; in this case, we extend the averaging period by a variable
amount up to $10^9$ years so that we can resolve star formation rates
down to $\sim 10^{-4} \mathrm{M}_{\odot} \yr^{-1}$.

The SFRs of our individual haloes are shown in Figure \ref{fig:sfr};
the horizontal dotted line shows the star formation rate of an
$L^{\star}$ LBG galaxy. This is derived from
\cite{2007arXiv0706.4091R}, wherein is stated
$M^{\star}_{\mathrm{AB}}(1700 \AA) = -20.8$. We adopt the conversion
ratio of \cite{1998ARA&A..36..189K}, but first correct the luminosity
for dust attenuation by a factor of $4.5$, which is the mean
correction given \cite[][ section 8.5]{2007arXiv0706.4091R}. We then
divide the final result by $1.6$ in order to consistently use the
Kroupa IMF (Section \ref{sec:simulations}), for which a larger number
of massive, hot stars are formed relative to the Salpeter IMF assumed
by \cite{1998ARA&A..36..189K}. This gives a final characteristic LBG
star formation rate of $39\, \msol\, \yr^{-1}$.  

We also plot a shaded band indicating the approximate SFRs obtained
for DLAs using the \cii technique \citep{2003ApJ...593..215W}. This is
intended to serve as a guide only -- there are a number of
uncertainties in the position of this band, including the intrinsic
complexity of the observations and the conversion from a star
formation rate per unit area (we have assumed a DLA cross-section of
$1 < \sigma/\kpc^2 < 100$, which is our approximate range for the most
common DLA haloes -- see Figures \ref{fig:xsec-raw} and
\ref{fig:xsec-hmf}).

By evaluating
\begin{equation}
\dot{\rho}_{\star} = \int \dd M_{\mathrm{vir}} f(M_{\mathrm{vir}}) \dot{M}_{\star}(M_{\mathrm{vir}})\label{eq:sfr-cosmo}
\end{equation}
(using a binning technique to discretize the integral) we derive a
global star formation rate at $z=3$ of $\sim 0.2 \msol \yr^{-1}
\Mpc^{-3}$. The UV dust-corrected and IR estimates in
\cite{2007arXiv0706.4091R} place this value between $0.05$ and $0.2
\msol \yr^{-1} \Mpc^{-3}$, but these are again sensitive to the IMF,
assuming a Salpeter form for their main results: for a Kroupa IMF one
should again reduce the rate by $\sim 1.6$.  This caveat should be
borne in mind, but overall our results are not unreasonable and a
detailed understanding of remaining discrepancies is beyond the scope
of the present work.

More importantly for our purposes, the star formation rate roughly
scales as $\dot{M}_{\star} \propto M_{\mathrm{vir}}^{1.4}$ (see fit in
Figure \ref{fig:sfr}). Given the low mass end of the halo mass
function is an approximate power law, $M_{\mathrm{vir}}^{-0.9}$, the
overall dependence of the integrand of equation~(\ref{eq:sfr-cosmo})
is rather shallow and a wide range of halo masses contribute to the
global star formation rate. Thus, there is no inconsistency in the
view that a typical ($M_{\mathrm{vir}} \sim 10^{10} \msol$) DLA has a
low star formation rate but that the DLA cross-section as a whole
contributes significantly to the star formation, and is largely
converted into stars by $z=0$.

We generate a cosmological DLA cross-section weighted sample of these
star formation rates (Section \ref{sec:cosm-conv}).  Figure
\ref{fig:dla-weighted-sfr} shows the resulting distribution, $\dd^2
N/(\dd X \dd \log_{10} \dot{M}_{\mathrm{star}})$. As expected by the
range of halo masses contributing to the DLA cross-section (Figure
\ref{fig:xsec-raw}), the star formation rate in most DLAs is much
lower than that in visible LBGs. Qualitatively, the simulated rate of
DLA star formation is consistent with or fractionally lower than the
observed star formation rate estimates from the \ciis~technique (see
above). Since we have not included upper limits in our approximate
observational band, this is an acceptable result.

\begin{figure}
\includegraphics[width=0.48\textwidth]{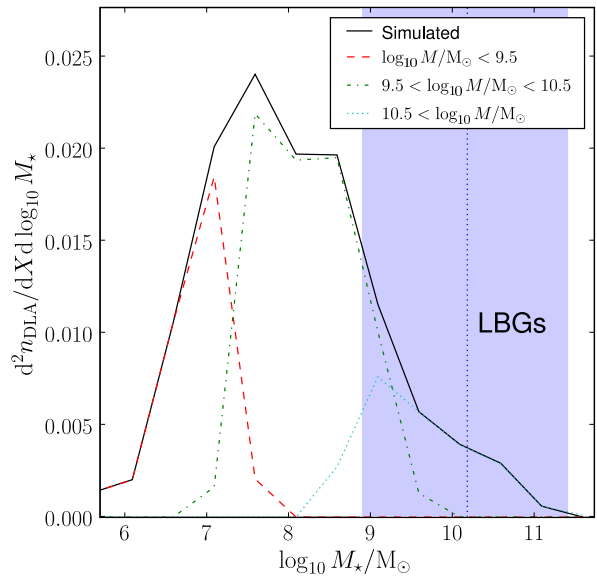}
\caption{The distribution of stellar masses of our DLAs. The vertical
  band and vertical dotted line show respectively the range of and
  median value for observed LBG stellar mass from
  \protect\cite{2001ApJ...562...95S} (adapted for consistency with our
  IMF). As expected, most DLAs have formed a relatively small
  population of stars, consistent with their low
  metallicities. }\label{fig:dla-weighted-smass}
\end{figure}

Finally, we investigated the total stellar mass accrued in our DLAs
(the integral of the star formation rate). This quantity can be
estimated in surveys of LBG by fitting their spectral energy
distribution deduced from multi-band photometry. In Figure
\ref{fig:dla-weighted-smass} we have plotted the distribution of our
DLA stellar masses and compared it with the estimated range of
observed LBG's stellar masses, using the results of
\cite{2001ApJ...562...95S} adapted as described above for our Kroupa
IMF. Most DLAs have formed a relatively small population of stars (the
distribution ranges over $10^{6}<M_{\star}/\msol<10^{11}$ but peaks
strongly at $M_{\star} \sim 10^{7.5} \msol$), consistent with their
low metallicities.  There is a strong dependency on the underlying
halo mass, which we have shown by plotting the contribution from
different halo mass ranges. This arises not only because the baryonic
mass rises roughly linearly with the virial mass, but also because the
star formation in low mass haloes is substantially suppressed by
feedback (see also Sections \ref{sec:metallicity} and
\ref{sec:feedback}).

The simulated $z=3$ DLAs are clearly being consumed very slowly, and
it will be a matter of considerable interest to understand the
ultimate fate of the gas, especially given the apparent realism of the
galaxies at $z=0$. We intend to address this question fully in future
work; for the present, we note that over 80\% of the neutral \hi~in
the major progenitor at $z=3$ has formed stars in our Milky Way type
galaxy (box MW) by $z=0$. However, this accounts for only 10\% of the
total stars; 10\% of the stars were in fact already formed at $z=3$,
with the remainder coming from accreted gas and stars in satellites.

\section{Consistency Checks and Effects of Parameters}\label{sec:consistency-checks}
\begin{table}
\begin{tabular}{lllll}
\hline
Brief Tag & $\langle M_{\mathrm{p,gas}} \rangle$ & $ \langle M_{\mathrm{p,DM}} \rangle$ & $\epsilon / \kpc$ & Comment \\
\hline
\textbf{MW} & $10^{5.2}\,\msol$ & $10^{6.2}\,\msol$ & $0.31$ & As Table \ref{tab:simulations} \\
\hspace{2mm}.LR &  $10^{6.9}$ & $10^{6.7}$ & $0.63$ & Lower Resolution \\
\hspace{2mm}.LR.ThF & $10^{6.9}$ & $10^{6.7}$ & $0.63$ & Thermal Feedback \\
\hspace{2mm}.LR.SS & $10^{6.9}$ & $10^{6.7}$ & $0.63$ & Self-Shielding \\
\hline
\textbf{Large} & $10^{6.1}$ & $10^{7.0}$ & $0.53$ & As Table \ref{tab:simulations}\\
\hspace{2mm}.LR & $10^{7.0}$ & $10^{7.9}$ & $1.33$ & Lower Resolution \\
\hspace{2mm}.LR.ThF & $10^{7.0}$ & $10^{7.9}$ & $1.33$  &Thermal Feedback \\
\hspace{2mm}.SS & $10^{6.1}$ & $10^{7.0}$ & $0.53$ & Self-Shielding \\
\hline
\textbf{Cosmo} & $10^{6.7}$ & $10^{7.6}$ & $1.00$  & As Table \ref{tab:simulations} \\
\hspace{2mm}.SS & $10^{6.7}$ & $10^{7.6}$ & $1.00$ & Self-Shielding \\
\hline
\end{tabular}
\caption{A summary of the comparison simulations used.}
\label{tab:comp-simulations}
\end{table}

In this section, we will discuss the effect of changing some details
of our simulations and processing, including the resolution and star
formation feedback prescription parameters. Since the effects are
rather minor overall, some readers may prefer to skip directly to our
conclusions (Section \ref{sec:discussion}). The extra simulations used
for the discussions below are summarised in Table
\ref{tab:comp-simulations}.

\subsection{Resolution and Star Formation Feedback}\label{sec:resolution}\label{sec:feedback}

\begin{figure}
\includegraphics[width=0.48\textwidth]{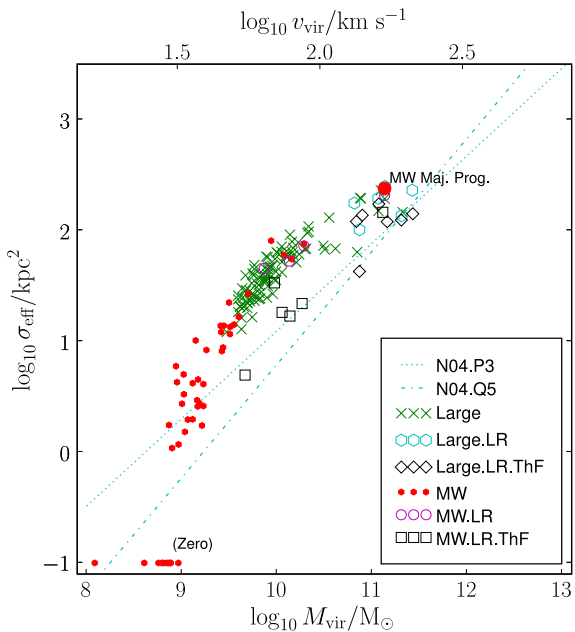}
\caption{As Figure \ref{fig:xsec-raw}, except we now plot only two
  simulations (MW, Large; dots and crosses respectively) and variants
  thereon -- see Table \ref{tab:comp-simulations} for a description of
  these.}
\label{fig:comparison-xsec}
\end{figure}

In this section, we describe a further suite of simulations which
probe the dependence of our overall conclusions on resolution and
feedback strength. Previous studies of DLAs have
been shown to be somewhat sensitive to the resolution
\cite[e.g.][]{2004MNRAS.348..421N,Razoumov:2005bh} and the main
difference between our simulations and previous work lies in the
feedback implementation, so that both these considerations are worth
exploring.

We described in Section \ref{sec:dla-properties-haloes} our resolution
criteria, demanding both that $N_{\mathrm{DM}}>1\,000$ and
$N_{\mathrm{gas}}>200$.  This is quite a conservative limit, based on
rerunning two of our simulations (``MW'' and ``Large'') at lower
resolution (for details see Table \ref{tab:comp-simulations}). For
haloes with coarser resolution than this stated limit, we found that
the DLAs started to exhibit scatter about the locus defined by their
higher resolution counterpart, and a slight tendency to underestimate
the DLA cross-section (while overestimating individual column
densities). These resolution effects are relatively mild compared to
some previously reported effects (e.g. \cite{1997ApJ...484...31G},
\cite{2004MNRAS.348..421N}, \cite{Razoumov:2005bh}). We suggest that
this is because our treatment of feedback (see Section
\ref{sec:simulations}), which leads to efficient energy deposition
from supernovae in dense areas, helps to impose an effective
scale-length below which the particles composing the ISM do not
collapse further.  Ultimately, given the impossible task of
maintaining sufficient dynamic resolution to track the actual gas
components making up the ISM, this is not an unreasonable
behaviour. Figure \ref{fig:comparison-xsec} shows the ``MW'' and
``Large'' simulations along with their low resolution counterparts
(``MW.LR'', ``Large.LR'') and demonstrates that the simulations are in
good agreement when restricted according to our resolution
criteria. The velocity widths from these low resolution simulations
were also overall in agreement with their high resolution
counterparts; however, as previously noted, the star formation
histories and metallicities only converge for a more stringent
resolution cut (Section \ref{sec:metallicity}).

We took the low resolution initial conditions and reran the
simulations, replacing our normal feedback implementation with a
purely thermal approach (i.e. the same energy per supernova was
produced, but the cooling switch-off was not implemented). This has
the well-known effect of causing the supernovae energy to be radiated
away over much less than the dynamical timescale
\citep{1992ApJ...391..502K,2000ApJ...545..728T} and is thus the
weakest conceivable mechanism. It is almost certain to be unphysical,
since the fast cooling times arise from the combination of high
temperature and density. This is caused by averaging properties of
unresolved regions; in fact the high temperature (blast interior) and
high density (undisturbed exterior) regions are presumably separated
in the true parsec-scale ISM.

Our cross-sections arising from this approach lie considerably closer
to those of \cite{2004MNRAS.348..421N} (Figure
\ref{fig:comparison-xsec}). We also find more usual results for
quantities such as the metallicity; because our metallicities converge
only at high resolutions (see Section \ref{sec:metallicity})
quantitative comparisons between our low resolution runs are a little
dangerous, but we see a very much weaker mass-metallicity relation
\citep[see also][]{2007ApJ...655L..17B}, with metallicities all
approximately $Z_{\odot}/5$. This is, as expected, in poor agreement
with observations but in better agreement with older
simulations. Recall that our fiducial feedback formulation causes a
reduction in metallicities primarily by suppressing star formation
efficiently, rather than causing any bulk outflows
\citep{2007ApJ...655L..17B}.

We also note that the velocity widths arising from particular haloes in
our thermal feedback simulations are comparable to, or somewhat ($\sim
5 \%$) {\it higher} than, the velocity widths in our fiducial
simulations. (The velocity widths agree between low resolution and
standard runs.) This does not result in a closer matching of the
cosmological velocity width distribution, because the proportion of
intermediate and high mass haloes is reduced.  We interpret this as
suggesting that the origin of our velocity widths is chiefly
gravitational, and that the thermal feedback runs form denser, compact
\hi~regions. This is no surprise: there is no physics in our
simulations that can plausibly give rise to large-scale bulk galactic
winds. Future improvement in DLA velocity width modeling may therefore
arise from an understanding of what drives such flows, and whether
cold gas can indeed exist within them.  However this understanding
will augment, not replace, a sufficient feedback model for the local
suppression of star formation; see Section \ref{sec:velocity-profiles}.

\subsection{Si\,II and low-ion regions}\label{sec:where-is-siii}

\begin{figure}
\includegraphics[width=0.48\textwidth]{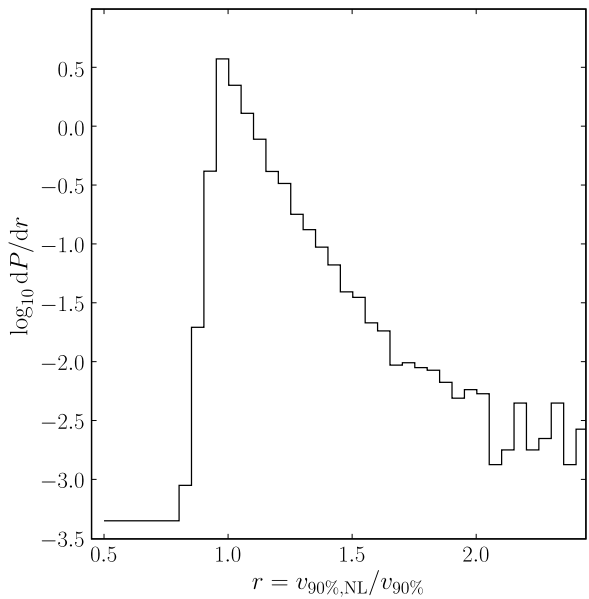}
\caption{The probability distribution of the ratio $r$ of the velocity
  width from the {\sc Cloudy}-based \siii~ ionisation runs to the
  original velocity widths which assume $n(\textrm{Si\,{\sc
      ii}})/n(\textrm{H\,{\sc i}})=n(\textrm{Si})/n(\textrm{H})$.
  Although higher velocity widths do arise by considering these
  effects, the mean differences are not large enough to have a
  significant impact on, for example, the overall distribution of
  observed DLA line widths (Figure \ref{fig:veldistrib-std}). }
\label{fig:cloudy-siii}
\end{figure}

One possible explanation for the underestimated incidence of high
velocity width absorbers in our simulations (Section
\ref{sec:veloc-width-distr}) is a misidentification of the precise
regions responsible for producing the low-ion profiles.  Therefore, we
briefly investigated the effect of relaxing our assumptions (Section
\ref{sec:self-shielding}) that $n(\textrm{Si\,{\sc
    ii}})/n(\textrm{H\,{\sc i}})=n(\textrm{Si})/n(\textrm{H})$.  This
assumption is based on comparing the energetics of the ion
transitions; however there is no actual physical mechanism coupling
the \siii~stage to \hi. This is to be contrasted with O{\sc i}, which
is strongly coupled to \hi~via a charge transfer mechanism
\citep[e.g.][]{2006agna.book.....O}. There are clear examples of DLAs
in which the O{\sc i} width is significantly smaller than the
\siii~width \cite[see, for example, the lowest two panels in Figure 6
of][]{1998A&A...337...51L} and thus there must be contributions to the
\siii~velocity structure from outside the cold \hi~gas.

Because DLAs are low metallicity environments (both observationally
and in our simulations), a first approximation to the silicon
ionisation problem can be achieved by taking the radiation intensity
from our hydrogen/helium radiative transfer code (Section
\ref{sec:self-shielding}) and calculating the equilibrium state of
Silicon. To address this in a simple way, we calculated a grid of {\sc
  Cloudy}\footnote{Version 07.02.01, available from {\tt
    www.nublado.org} and last described by \cite{1998PASP..110..761F}}
models which varied in the local density, temperature and incident
radiation strength indexed by the single photoionisation parameter
$\Gamma_{\mathrm{HI}+\gamma \to \mathrm{HII}+e}$. For each particle in
our simulation, we calculated the value of Si{\sc ii}/Si$(\Gamma, T,
\rho)$ by interpolating these models. Then we recalculated our DLA
sightlines (Section \ref{sec:dla-properties-haloes}) using the new
Si{\sc ii} values, rather than the old assumed values. We used exactly
the same set of offsets and angles, so that the final
reprocessed-catalogue can be compared on a per-sightline basis. Our
sightlines extend through the entire box so that any trace gas outside
our haloes could theoretically contribute to the velocity
width. However, the combination of the decreasing neutral gas density
and decreasing metallicities makes the intergalactic contribution to
the unsaturated line widths extremely minor.

We have plotted, for all sightlines, the probability distribution of
the ratio $r$ of the updated velocity widths to the original in Figure
\ref{fig:cloudy-siii}.  A number of sightlines did, after
reprocessing, display significantly higher velocity widths (as much as
doubling in some cases). Thus for individual systems, these
corrections can be important.  However (noting that the $y$-axis in
the plot is logarithmic) it is clear that the significantly increased
velocity width systems are too rare to make a significant impact on
the distribution of velocity widths (Figure \ref{fig:veldistrib-std}),
a fact we verified explicitly.

\subsection{Cosmological Model}\label{sec:cosmological-model}

The parameters used throughout this work were fixed when running the
first of the simulations described. Since then, our knowledge of
cosmological parameters has improved thanks to the expanding wealth of
constraints from galaxy surveys measuring the baryon acoustic
oscillations (BAO), supernovae surveys and most significantly three-
and five-year WMAP results
\citep{2006astro.ph..3449S,2008arXiv0803.0586D}. Our chosen parameters
are $(\Omega_{\mathrm{M}}, \Omega_{\mathrm{b}},
\Omega_{\mathrm{\Lambda}}, \sigma_8, h, n_s)= (0.30, 0.044, 0.70,
0.90, 0.72, 1.0)$, whereas the latest WMAP, BAO and SN results
combined suggest $(\Omega_{\mathrm{M}}, \Omega_{\mathrm{b}},
\Omega_{\mathrm{\Lambda}}, \sigma_8, h, n_s)= (0.269, 0.046, 0.721,
0.82, 0.70, 0.96)$. It is natural to question whether such differences
can have a significant effect on our results.

Given our confined cross-section (Section \ref{sec:halo-properties}),
any effect can be split into two components: the change in the halo
mass function, and the change in the individual objects. Considering
the halo mass function first, since at $z=3$ the relative density of
$\Lambda$ to CDM is suppressed by a factor of $4^3=64$, the difference
in $\Omega_{\Lambda}$ has a negligible effect. Further, the value of
$\Omega_b/\Omega_{\mathrm{M}}$ has only minor consequences for the
transfer function (amounting to the form of the acoustic
oscillations). The most important difference, therefore, should arise
from the reduction of $\sigma_8$: this will decrease the scale on
which fluctuations have become non-linear, and hence reduce the number
of high mass objects.  However, we also need to consider the lower
value of $n_s$. The high mass end of the halo mass function is less
affected (since it is closer to the normalization scale), but at the
low mass end (corresponding to high $k$) there are a lower than
expected number of haloes. Overall then, the effect of updating our
halo mass function is to slightly reduce the line density of DLAs (to
$l_{\mathrm{DLA}} = 0.049$, which is marginally lower than the
observed line density, but acceptable given the uncertainties in our
simulations) and shift the cross-section to somewhat lower
masses. However, we verified that this had negligible effect on, for
example, our velocity width distribution. This is because,
fortuitously, the two effects produce an almost flat value of
$f_{\mathrm{WMAP}}(M)/f_{\mathrm{fiducial}}(M) \simeq 0.7$ over
$10^8<M/\mathrm{M}_{\odot}<10^{11}$. 

On halo scales, the $\sim 5 \%$ shift in the value of $\Omega_b$ must
be the dominant effect. We do not believe this will have a large
effect on our simulations, being a small correction, especially since
equilibrium considerations determine the physics on these scales. A
complete resolution of this question, however, awaits new simulations
with the updated values.

\subsection{Radiative Transfer}\label{sec:radiative-transfer}

We have incorporated a correction for self-shielding (Section
\ref{sec:self-shielding}) which is coarse compared to some recent
simulations \citep{Razoumov:2005bh,2007arXiv0710.4137R}. On the other
hand, the complexity of the fine-structure of the ISM today (and
presumably also at $z=3$) means that even the most advanced
simulations cannot capture its ``microscopic'' (i.e. parsec scale)
behaviour. \cite{2004MNRAS.348..421N} employed a sub-resolution
two-phase pressure equilibrium model for the ISM
\citep{2003MNRAS.339..289S,1977ApJ...218..148M}, apparently obviating
the need for radiative transfer since the cold clouds are assumed to
be fully self-shielded and the warm ambient medium to be optically
thin. This is a promising approach, but it is not entirely clear how
it links to the messy observational picture in which atomic gas exists
in distinct warm and cold phases, ionized gas in warm and hot phases,
and photoionized regions occur within the cold phase around star
formation (\hii~regions) as well as in the diffuse disk ISM.

For this work, our essential aim was to identify large scale regions
in which the cosmological UV background should be significantly
suppressed. We verified our code using a variety of tests comprising
comparisons of equilibrium ionisation front positions with simple
constant temperature \textsc{Cloudy} models. While our radiative transfer code
performs well in simple plane-parallel setups, it resolves only 6
angular elements and its true 3D performance is therefore
significantly degraded. However, we believe that for the purposes of
the present study it is adequate: in both optically thin and optically
thick limits, the angular resolution is of little importance; only in
the transition regions will significant differences arise. We ran a
final test in which we post-processed our ``Large'' box after rotating
it through 45$^{\circ}$, thus providing a different set of angular
elements to the code.  None of our DLA results were significantly
affected by this change.

We do not include the UV emission of star forming regions in our
calculation. This would introduce a significant extra complexity to
our algorithm without any real benefits for the following reason. The
star formation rate of most DLAs is known observationally -- and in
our simulations -- to be $\lsim 1 \msol \yr^{-1}$ (see Section
\ref{sec:star-formation-rates}, Figure \ref{fig:dla-weighted-sfr}).
For such low star formation rates the local contribution to the UVB is
comparable to, or usually significantly smaller than, the $z=3$ field.
This can be estimated, for example, using the data of
\citeauthor{1987ASSL..134..731B} 1987, or more usually is obtained
directly from observational data -- see Introduction. Although the
theoretical picture of the importance of local sources is not
straightforward
\citep[e.g.][]{2005ApJ...620L..91M,2006ApJ...643...59S}, we are
confident that these direct observational upper limits justify our
approximation.

There is another slight inconsistency in our approach, in that the
radiative transfer post-processor identifies certain regions of gas as
neutral which in the live simulations is treated as ionized; further,
UV heating occurs for the uniform background even in regions which are
later identified as self-shielding. To assess the magnitude of this
effect, we re-simulated ``MW.LR'', ``Large'' and ``Cosmo'' using a
local UVB attenuation approximation (``SS'' simulations in Table
\ref{tab:comp-simulations}). In these simulations, for each gas
particle the strength of the UVB used for ionization equilibrium and
heating calculations was reduced by the mean attenuation for particles
of that density in the post-processed simulation.  We continue to
apply our radiative transfer post-processor, but the corrections
involved are smaller in the new outputs.

\begin{figure}
\includegraphics[width=0.48\textwidth]{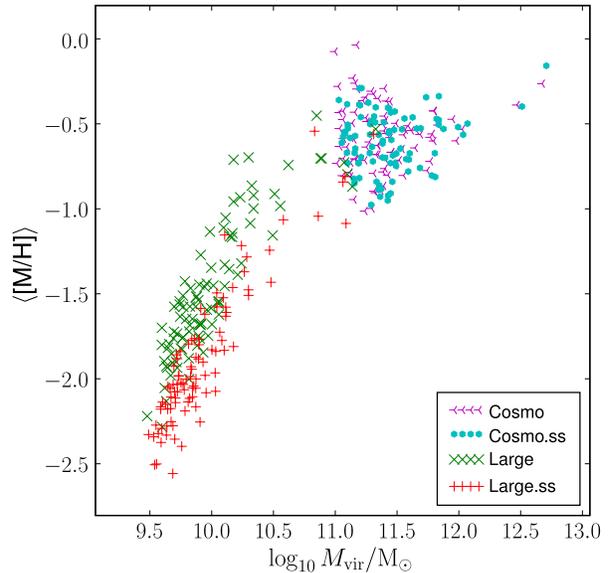}
\caption{The mean (sight-line weighted) metallicity of DLAs in haloes
  from the ``Cosmo'' \& ``Large'' boxes (tripod and cross symbols
  respectively) contrasted with their live self-shielding counterparts
  (dots and plus symbols). The most significant effect of including
  live self-shielding, other than a slight overall increase in DLA
  cross-sections, is a decrease in star formation rates in low mass
  haloes, causing them to display systematically lower
  metallicities. For clarity in this plot, we have only displayed the
  first 100 haloes from the ``Cosmo'' boxes.}
\label{fig:ss-metallicities}
\end{figure}

In general, the differences between the ``live'' self-shielding and
original runs were rather minor. They consisted of a slight increase
in the DLA cross-section ($\sim 0.2$ dex) and a reduction in the star
formation rates. Consequently the cold gas metallicity of these low
mass haloes was seen to drop by up to $0.5$ dex (Figure
\ref{fig:ss-metallicities}) -- this was apparently the biggest change
caused. In fact, this could help to resolve the slight paucity of low
metallicity ([M/H]$<-2.0$) DLAs in our simulations (Figure
\ref{fig:metals}), although without a fuller set of statistics we were
not able to verify this directly. Importantly, however, we reproduced
all our measured distributions replacing ``Cosmo'' with ``Cosmo.SS''
and ``Large'' with ``Large.SS''. The effects of this were dominated by
an increase in overall line density to $l_{\mathrm{DLA}} = 0.080$ (see
Section \ref{sec:column-dens-distr}). This takes us further away from
the observed result $l_{\mathrm{DLA}}=0.065 \pm 0.005$. However,
excepting normalization, the distribution of properties was left
almost unchanged. We believe these effects warrant further
investigation in the future, but their direct effects seem to be
fortuitously small.  One should be aware that, as well as the
self-shielding effect which reduces the UV heating and hence
equilibrium temperature and pressure in the ISM, other inaccuracies in
the detail of the ISM (such as magnetic pressure) could easily
contribute opposite effects of similar magnitudes; however their study
is well beyond the scope of this work.

\subsection{Intergalactic DLAs?}\label{sec:no-interg-dlas}

It is clear from Figure \ref{fig:projected-hi} that our DLA
cross-section is confined to within the virial radii of our host
haloes. This is in qualitative agreement with the majority of previous
DLA simulations listed in Table \ref{tab:previous-simulations}, and is
an assumption of all DLA semi-analytic models that we are aware
of. However, it contrasts with the recent results of
\cite{Razoumov:2005bh} and \cite{2007arXiv0710.4137R} (R06 and R08
respectively) in which as much as 50\% of the DLA cross-section
resides outside the virial radii of haloes (see R08 Table 2 and the top
right panel of R06 Figure 3).

Although rough estimates suggest that DLAs should form in dark matter
haloes, it is not unreasonable for DLAs to form in overdense
surrounding regions. Given the requirement for self-shielding,
$n_{\mathrm{HI}} \gsim 10^{-2} \cm^{-3}$ \citep{1998ApJ...495..647H},
one may compare the gravitational mass required to confine such gas
($M_{\mathrm{grav}} \sim N_{\mathrm{HI}} kT / G m_p n_{\mathrm{HI}}$)
to the total gas mass enclosed in the DLA ($\sim m_p
n_{\mathrm{HI}}^{-2} N_{\mathrm{HI}}^3$), giving the ratio
\begin{eqnarray}
\frac{M_{\mathrm{grav}}}{M_{\mathrm{gas}}} \sim 2 \left(\frac{T}{10^4 \mathrm{K}}\right)\left(\frac{N_{\mathrm{HI}}}{10^{20.3} \cm^{-2}}\right)^{-2} \left(\frac{n_{\mathrm{HI}}}{10^{-2} \cm^{-3}}\right) \textrm{.}
\end{eqnarray}
Thus, for DLAs with column densities close to the lower limit, it may
be possible for the self-gravity of gas, or slight dark matter
overdensities, to obviate the need for a fully collapsed dark matter
halo. This is especially true if (unlike in our simulations) gas is
allowed to cool efficiently to temperatures $\ll 10^4
\mathrm{K}$. 

The bigger problem is in understanding how the gas cools to become
neutral in the first place. One key difference between our simulations
and those of R06, R08 is that the latter include an approximate
algorithm to correct the temperature for shielding effects, whereas we
keep our temperatures constant during the radiative transfer
processing (Section \ref{sec:self-shielding}). While our approach is
clearly an approximation, the latter may overcool gas since dynamical
and feedback heating effects could easily become significant as the UV
field drops. The only way to correctly resolve this problem is (as in
R06, but not R08) to include radiative transfer in the live
simulations.

It will be interesting to see how this issue and a physical
understanding of it develop with future generations of simulations and
radiative transfer codes. But while it remains to be adequately
resolved, we should note that ({\it a}) the effect in R06/08 is
resolution dependent, and therefore the physical status is not
transparent; ({\it b}) our self-shielding runs suggested the effect
was rather small (Section \ref{sec:radiative-transfer}), at least with
the local approximation and ({\it c}) we have produced a DLA
cross-section which matches nearly all observational constraints,
which lends some indirect reinforcement to the validity of our
approximation.

\section{Discussion and Conclusions}\label{sec:discussion}\label{sec:limit-future-direct}

\subsection{Overview}

We have investigated the occurrence of DLAs in a series of simulations
(see Governato et al. \citeyear{2007MNRAS.374.1479G},
\citeyear{2008arXiv0801.1707G} and \citeauthor{2007ApJ...655L..17B} 2007 -- in
this text referred to as G07, G08 and B07) which produce galaxies at
$z=0$ with as near as currently possible to realistic physical
properties (see Introduction). These high resolution simulations
include a physically motivated star formation feedback prescription
with only one free parameter (the efficiency of energy deposition)
which is set by observations of low redshift star formation. Thus, as
well as providing information on DLAs themselves, this work is an
independent cross-check of the formation process of the G07/B07/G08
galaxies.

To produce cosmological statistics from a smorgasbord of boxes, we
used a non-parametric weighting process, in effect correcting the halo
mass function of the combined sample. This does not involve any
fitting or parametrization, and hence no extrapolation: all the
results presented in this paper are derived directly from resolved
regions of simulations.

The picture that emerges from our simulations is of a cosmological DLA
cross-section predominantly provided by intermediate mass haloes,
$10^{9}<M_{\mathrm{vir}}/\mathrm{M}_{\odot}<10^{11}$. These ranges are
somewhat higher than many early simulations suggested
\cite[e.g.][]{1997ApJ...486...42G,2001ApJ...559..131G}, and are close
to the range of $10^{9.7}\lsim M_{\mathrm{vir}}/\msol \lsim 10^{12}$
suggested by observations of DLA/LBG correlations
\citep{2006ApJ...652..994C}. Our DLAs form stars at low rates
(predominantly $\lsim 0.1 \msol\, \mathrm{yr}^{-1}$) and have
metallicities in the range $-2.5<\mathrm{[M/H]}<0$ with a median of
approximately $Z_{\odot}/20$, both in agreement with observations.  We
also investigated the distribution of total stellar masses of DLAs,
finding them to be predominantly spread over $10^6 <
M_{\star}/\mathrm{M}_{\odot} < 10^{10}$, with a peak at $\sim 10^{7.5}
\msol$.  By $z=0$ the majority of the neutral gas forming the DLAs has
been converted into stars, but during this time substantial merging
complicates the direct identification of low redshift galaxies with
high redshift absorbers.  In plots showing overall halo properties we
have marked the location of the major progenitors to our $z=0$ Milky
Way like and dwarf-type galaxies (from boxes ``MW'' and ``Dwf''
respectively); they appear to have been fairly typical DLAs.

\subsection{Mass-Metallicity Relationships}

A tight relationship between dark matter halo mass and star formation
rate (Figure \ref{fig:sfr}) underlies a strong mass-metallicity
relation; in Figure \ref{fig:velmetals} we have shown that this is
manifested by a correlation between DLA velocity widths and
metallicities \citep[for equivalent observational results
see][]{2006A&A...457...71L,2008ApJ...672...59P}. Further, the same
simulations produce a realistic mass-metallicity relation for $0<z<2$
galaxies (B07), suggesting that these simulations capture the metal
enrichment of collapsed systems in a meaningful way.  The DLA result
is not significantly affected by gradients within the disks of our
forming galaxies, but the DLA sightlines do preferentially sample
regions of gas with {\it higher} metallicities (by factors $\sim 2$)
than the mass-weighted halo mean.  Further work on metallicities in
these simulations is ongoing, in particular regarding the importance
of metal diffusion -- the metals are simply advected with the SPH
particles, which can lead to inaccuracies where spatial gradients are
important.  In connection with this work, it will be interesting to
see whether the metal enrichment of the simulations' less overdense
regions also appears realistic, providing a connection to the
importance (or otherwise) of outflows (Section
\ref{sec:velocity-profiles}).

Recent work has suggested that the origin of the observed relationship
between kinematics as measured by Mg\,\textsc{ii} rest-frame
equivalent widths ($W_r^{\lambda 2796}$) and metallicity of DLAs
should not be relied upon as an indicator of an underlying variation
with mass \citep{2008arXiv0803.3944B}. However, this claim is not
directly relevant to fiducial $v_{90 \%}$ velocity width
determinations which are measured from unsaturated or mildly saturated
transitions; MgII is a strongly saturated absorption line, and as such
$W_r^{\lambda 2796}$ can easily be affected by trace gas in the outer
regions of galaxies (or perhaps even the ambient IGM).

\subsection{Velocity Profiles}\label{sec:velocity-profiles}

Overall, we reproduce the spread of velocity widths seen in DLA
systems (Figure \ref{fig:veldistrib-std}) and approximately match the
distribution of low velocity width systems ($v<100\, \kms$).  However
the long standing difficulty of underestimating the observed incidence
rate of high velocity widths ($v>100 \, \kms$) persists in our
simulations -- although the discrepancy is rather small compared with
older simulations and comparable to that seen in the recent work by
\cite{2007arXiv0710.4137R}. It is not entirely clear why simulations
in general have encountered such persistent difficulty in this
respect. There are essentially three possibilities:

Firstly, it is possible that our assumption $n(\textrm{Si\,{\sc
    ii}})/n(\textrm{H\,{\sc i}})=n(\textrm{Si})/n(\textrm{H})$
oversimplifies the identification of regions responsible for the
low-ion profiles and leads to systematic underestimates of the
velocity dispersion.  We investigated this avenue briefly (Section
\ref{sec:where-is-siii}), but the correction did not appear sufficient
to resolve the discrepancy -- although this could be sensitive to the
description of the ISM on sub-resolution scales (see also below). We
ran a brief test in which {\it all} gas was assumed to contribute to
the velocity width -- in this (completely artificial) scenario, the
cosmological rate of high velocity widths is {\it over}-estimated,
showing that the required motions are, at least, ``available'' in this
sense.

Secondly, perhaps the internal gas kinematics still require some
correction. Although we did not, for this work, attempt a rigorous
decomposition, qualitative inspection suggested that a {\it mixture}
of disk kinematics \citep{1998ApJ...507..113P} and merging clumps
\citep{1998ApJ...495..647H} are responsible for the final spread of
velocities. We did not find that our feedback made any significant
contribution via turbulent motions to the velocity widths (Section
\ref{sec:feedback}) -- hardly a surprise, given that the coupling to
the ISM is achieved entirely thermally. A better understanding of
feedback is likely to help the kinematical situation by inducing bulk
outflows.  It would be interesting to see how the simulations of
\cite{2004MNRAS.348..421N} perform in reproducing DLA line profiles,
since they include a phenomenological model of galactic winds.
Promising progress in placing such winds on a more physical footing was
recently described by \cite{2007arXiv0712.3285C}, whose simulations
reproduce such bulk motions starting from an explicit model for the
ISM.

Thirdly, it is possible that our cross-section is associated with, on
average, haloes of insufficient mass; in other words, DLAs at the high
mass end could be overly compact. This in turn could be connected to
the more well-known angular momentum problem in which simulated disk
galaxies have underestimated scale lengths
\citep{2000ApJ...538..477N,2001ApJ...554..114E}.  Our simulations, as
was noted in the Introduction, go some significant way towards a
resolution of the issue for $z=0$ disk galaxies by combining
sufficient feedback \citep{2006astro.ph..2350S} with high resolution
\citep{2007MNRAS.375...53K}; our elevated DLA cross-sections for
intermediate-mass haloes are connected to this same feedback (Section
\ref{sec:resolution}). As the mechanisms preventing angular momentum
loss are further understood, these same mechanisms may continue to
increase the DLA cross-section for high mass haloes. On the other hand,
our matching of the metallicity distribution (Figure \ref{fig:metals})
provides an alternative joint constraint on the star formation history
and mass of responsible haloes -- and is in good agreement with
observations. Any change in the mass of haloes comprising our DLA
cross-section would therefore need to be compensated by a change in
our virial mass -- star formation relation (Figure~\ref{fig:sfr}).

\subsection{Column Density Distribution and Dust Biasing}\label{sec:column-dens-disc}

As we discussed in Section \ref{sec:column-dens-distr}, we match the
overall column density distribution well, but somewhat overestimate
the number of observed systems with $N_{\mathrm{HI}}>10^{21.5}
\cm^{-2}$. This causes us to slightly overpredict value of
$\Omega_{\mathrm{DLA}}$ (we obtain $\Omega_{\mathrm{DLA,sim}} = 1.0
\times 10^{-3}$, whereas the SDSS data yield
$\Omega_{\mathrm{DLA,obs}} = 0.8 \times 10^{-3}$). The simulated value
of $\Omega_{\mathrm{DLA,sim}}$ is obtained by imposing an upper
cut-off on the column density of systems contributing to the measured
$\Omega_{\mathrm{DLA}}$ because stronger systems are too rare to be
observed with current observational datasets. But in our simulations,
these same systems actually contribute significantly to the cosmic
density of DLAs; when included in the census, they boost the total to
$\Omega_{\mathrm{DLA,sim}}=1.4 \times 10^{-3}$. Thus in our picture,
``invisible'' systems contribute about $30 \%$ to the \hi~budget; but
functional fitting of the observed data from SDSS suggests a much
steeper cut-off at high column densities than our simulations, and
hence that the observed value $\Omega_{\mathrm{DLA,obs}}$ given above
should be essentially correct \citep{2005ApJ...635..123P}. While the
precise observational extrapolation is subjective and the fit is
driven by a couple of bins with $N_{\mathrm{HI}}>10^{21.5} \cm^{-2}$,
there is a strong constraint in the lack of systems at high column
density.  For instance, only three systems with
$N_{\mathrm{HI}}>10^{21.8} \cm^{-2}$ are observed in the SDSS DR5
survey, whereas our simulation predicts ten over the same path length
(a glance at the Poisson distribution function shows that this is a
significant discrepancy). 

It is notable that host galaxies of gamma ray bursts (GRBs), which
trace a sample weighted towards higher column densities, routinely
display column densities of $N_{\mathrm{HI}} \sim 10^{22} \cm^{-2}$
\citep{2006A&A...460L..13J} -- so that such systems certainly exist
but have a smaller cross-section than our simulations suggest. It is
possible that we overpredict the number of high column density systems
because our simulations are inadequate in the responsible regions,
which are cold and dense: one should form H$_2$ but the complex
physics responsible is not implemented in our
code. \cite{2001ApJ...562L..95S} suggested that this molecular cloud
formation could determine a characteristic cut-off for the column
density distribution. On the other hand, observations suggest that
DLAs harbour very little molecular gas, with fractions at the very
most $f_{\mathrm{H_2}} \sim 10^{-2}$ \citep{2003MNRAS.346..209L}. Of
course the typical DLA sightline could miss high density clumps within
giant molecular cloud formations, which would mean the global fraction
of H$_2$ could be somewhat higher.  It is therefore currently unclear
how feasible this explanation could be.

It has been claimed in the past \cite[e.g.][]{2003ApJ...598..741C}
that DLA dust obscuration effects should substantially reduce the
number of observed high $N_{\mathrm{HI}}$ systems relative to their
true abundance. However, this is hard to reconcile with the most
recent observations; in particular, radio-selected samples (which are
unlikely to be biased by dust considerations) show little if any
evidence for increased high $N_{\mathrm{HI}}$ absorbers
\citep{2001A&A...379..393E,2006ApJ...646..730J,Ellison08}. Further, in
a companion paper \citep{PontzenDLADust} we show that
there is only marginal statistical evidence for dust-induced
obscuration in the joint distribution of observed DLA column densities
and metallicities, which are in fact nearly uncorrelated.  This also
arises quite naturally in our simulations, because the dependence of
$N_{\mathrm{HI}}$ on the underlying halo mass is weak (Figure
\ref{fig:colden}), whereas metallicities and kinematics are strongly
correlated with the halo mass. Combining all the above considerations
with our successful reproduction of the observed distribution of
metallicities (Section \ref{sec:metallicity}), we believe dust-biasing
is unlikely to have a major part to play in a future understanding of
DLAs.

The very slight trend which is observed linking the column density to
the halo mass in our simulations biases low column density
observations in favour of finding low mass haloes (Figure
\ref{fig:colden}).  We have not explicitly studied the sub-DLA
population, so that these results are not directly comparable with the
work of \cite{2007A&A...464..487K}, in which it is suggested that
sub-DLAs preferentially probe more massive haloes than DLAs. However,
there is certainly a tension between these results which deserves some
attention in the future.

\subsection{Bimodality}\label{sec:bimodality}

Although we did not specifically set out to investigate the
possibility of bimodality within the DLA population
\citep{2008arXiv0802.3914W}, we saw a bifurcation in our cross
sections as a function of \hi~mass (Figure \ref{fig:xsec-vs-hi}).  The
bifurcating lower branch in Figure \ref{fig:xsec-vs-hi} has higher
virial masses for a given \hi~mass, wider velocity profiles, higher
star formation rates and higher mean temperatures. It is important to
emphasize that these systems, which are perhaps undergoing a
starbursting phase, account for only $\sim 2\%$ of the DLA
cross-section; our results are therefore not directly comparable to
the observational claim of two populations of roughly equal
importance.

We ran some careful tests to try to understand the origin of this
effect, but it essentially remains work in progress: further
simulations and detailed studies of the evolution of these objects
will be key to a full account. The critical halo mass,
$M_{\mathrm{vir}} \sim 10^{10.5} \msol$, is probably too small to be
directly related to the shock-stability model of
\cite{2006MNRAS.368....2D}. There are no spatial correlations which
would suggest the abnormal haloes are associated with protocluster
regions. We verified that the results from the ``Large'' box were
consistent with being drawn from a subvolume of the ``Cosmo''
box. Since the low cross-section branch objects account for such a
small fraction of our DLAs, we felt justified in postponing their full
study to later work.

\subsection{Future Work}

In general, we have commented throughout that the lack of resolution
on fine ISM scales (a common feature of all current cosmological
simulations, likely to persist into the foreseeable future) is a
fundamental limitation. The ISM is a complex mixture of gas in
different phases with variations on truly tiny scales. It is also
interesting to note that observations suggest that the pressure
contributed by magnetic effects in the disk of our galaxy and other
local galaxies is comparable in magnitude to turbulent kinetic
pressure \citep[e.g.][]{2005LNP...664..137H}; hence magnetic effects
could contribute at least as much as stellar feedback to an
understanding of the formation of disk galaxies, at least at $z=0$. It
is not impossible to fathom that magnetic fields could have an
important role to play in a better understanding of DLAs.  But
regardless of any particular physical processes, our feeling is that
long term future focus should remain on improving our understanding of
the coarse-grained behaviour of such a mixture, i.e. in developing
sub-grid physics models. This conclusion was reached independently by
\cite{2007arXiv0710.4137R}.

Even with our current set of simulations, it will be interesting to
investigate in more detail the predictions of these simulations with
regard to the LBG population; in Section
\ref{sec:star-formation-rates} we showed that the total star formation
rate at $z=3$ may be very slightly overpredicted compared to
observation. But, in work not presented here, we have also
investigated the shape of the luminosity function (at $z=3$ and higher
redshifts) and found both $L_{\star}$ and the faint end slope to be
consistent with observations. Remaining shortcomings could easily be
accounted for by details of the IMF, suggesting that the underlying
populations are quite reasonable (Governato et al, in prep).

In terms of the properties of DLAs, this work opens the door to an
array of further studies. In particular, we intend to address in more
detail the time evolution of our DLA population and the transfer of
gas through hot, warm and stellar phases in the near future. Given the
realistic nature of metals in our $z=3$ DLAs and in our galaxies at
$z=0$ \citep{2007ApJ...655L..17B}, the details of the intervening time
could, amongst other things, shed light on the ``missing metals''
problem \citep{2006astro.ph..3066P}. Further, it is known
observationally that the metallicities, velocity widths and column
density distributions evolve only slightly with decreasing redshift;
will our simulations reproduce such slow evolution, which depends on
finely balancing the cooling of gas, forming \hi, with the formation
of stars, destroying it? If so, the manner in which this is
accomplished will be of considerable interest. (For instance, do
individual objects maintain the same DLA cross-section, or is the slow
evolution only seen over cosmological averages?)  Although flawed in
many respects, simulations offer a guide to our understanding of these
issues which should not be ignored -- as long as it is taken with a
healthy pinch of salt.

\aponly{
\todo{Many previous simulations are sensitively
dependent on the resolution and it is not clear when to expect these
to converge \cite[e.g.][]{Razoumov:2005bh}; however, the inclusion of
a resolution-independent scheme for efficient stellar feedback
prevents unrealistic gas collapse and thus produces a natural cut-off
to the scales of interest in a coarse-grained description (this
process is better studied in the context of the scale lengths of the
final galaxies; see Introduction).}
\cite{Razoumov:2005bh,2007arXiv0710.4137R} observed that as their
resolution is increased, the velocity widths increase also as gas is
able to collapse to more condensed clumps. We do not see this effect
(see Section \ref{sec:resolution}), and again this can be explained by
the natural cut-off introduced by our resolution-independent treatment
of feedback.}

\section*{Acknowledgments}

We thank Jason X. Prochaska for providing observational data,
Francesco Haardt for UVB models and the referee, Peter Johansson, for
a thorough and helpful report. AP is supported by a STFC (formerly
PPARC) studentship and scholarship at St John's College, Cambridge and
gratefully acknowledges helpful conversations and communications with
George Efstathiou, Gary Ferland, Johan Fynbo, Alexei Razoumov, Emma
Ryan-Weber and Art Wolfe. FG acknowledges support from a Theodore
Dunham grant, HST GO-1125, NSF ITR grant PHY-0205413 (supporting TQ),
NSF grant AST-0607819 and NASA ATP NNX08AG84G.  Most simulations were
run at the Arctic Region Supercomputing Center. Part of the analysis
was performed on \textsc{Cosmos}, a UK-CCC facility supported by HEFCE
and STFC in cooperation with SGI/Intel utilising the Altix 3700
supercomputer.

\bibliographystyle{mn2e} {\small \bibliography{/home/app26/documents/refs}}

\appendix
\vspace{-0.5cm}
\section{SPH Calculations}\label{sec:sph-calculations}

For convenience we will denote the SPH averaging procedure by square
brackets, i.e. for any quantity $Q$
\begin{eqnarray}
  [Q](\vec{r}) & \equiv & \int \dd^3 r' W(\vec{r},\vec{r}'; h(\vec{r'})) Q(\vec{r}') \\
  & \simeq & \sum_i \frac{Q_i}{\rho_i} W(\vec{r},\vec{r}_i; h_i)
\end{eqnarray}
where $W$ is the smoothing kernel and the smoothing length for each
particle is given by $h_i$, which notionally corresponds to a
smoothing field $h(\vec{r})$. For each sight-line, the $z$ axis is
aligned such that the line of sight is parametrized by $\vec{r} =
(0,0,z)$. The column density is calculated as:

\begin{eqnarray}
N_{\himath} & = & \int_{-\infty}^{\infty} \dd z~ [n_{\himath}](\vec{r}) \\
& = & \sum_i \frac{n_{\himath,i} m_i h_i}{\rho_i} W_{\perp}(|\vec{r}_i|_{\perp};h_i) \label{eq:colden-sph-approx}\\
W_{\perp}(|\vec{r}_i|_{\perp};h_i) & \equiv & \frac{1}{h_i^2} \int_{-\infty}^{\infty} \dd z~ \mathcal{W}((\frac{x_i^2 + y_i^2}{h_i^2} + z^2)^{1/2}) \label{eq:proj-kern}
\end{eqnarray}
where $n_{\himath}(\vec{r})$ is the number density of \hi~ ions, and
$n_{\himath,i}$, $m_i$, $\rho_i$ refer to each SPH particle's \hi~
number density, mass and density respectively and we have assumed a
fiducial form for the kernel $W(\vec{r_1},\vec{r_2};h) =
\mathcal{W}(|\vec{r}_1-\vec{r}_2|/h)/h^3$. The projected kernel
(\ref{eq:proj-kern}) is calculated numerically; however, since
conventionally $\mathcal{W}$ has compact support, so too does
$W_{\perp}$ which need only be tabulated once for a finite range of
values. The upper limit depends on the choice of kernel, but is
conventionally $\max(d) = 2h$.  The sum (\ref{eq:colden-sph-approx})
is taken over all particles, although in practice many particles are
culled by considering their distance from the line of sight.

In general, by projecting the kernel in this way one obtains an
efficient and justifiable method for calculating any projected
quantity. Integrated quantities follow directly by replacing
$n_{\himath,i}$ in equation (\ref{eq:colden-sph-approx}) with the
appropriate particle property, subject to the following caveat: there
is an ambiguity in calculations of this type involving more than one
particle property or functions of particle properties. In particular
we will consider the generation of an absorption profile:
\begin{equation}
  \tau_{\mathrm{tot}}(\lambda) = \int_{-\infty}^{\infty} \dd z~ n(\vec{r}) \tau_{\mathrm{v}}\left(\lambda\left[1-v_z(\vec{r})/c\right];T(\vec{r})\right)\label{eq:tau-sph}
\end{equation}
where $\tau_{\mathrm{tot}}(\lambda)$ gives the total line of sight
optical depth for wavelength $\lambda$ and $\tau_{\mathrm{v}}$ gives
the absorption (Voigt) cross-section with parameters assumed for the
ion under consideration and temperature broadening $T(\vec{r})$. There
are (at least) two different methods for approximating this.

The first method is easiest to evaluate, and lumps the integrand into a
single quantity which is to be SPH-interpolated:
\begin{eqnarray}
\tau_{\mathrm{tot}}(\lambda) & = & \int_{-\infty}^{\infty} \dd z~ \left[ n \tau_{\mathrm{v}} \left(\lambda\left( 1-v_z/c\right) ;T \right) \right](\vec{r})  \\
& = &  \sum_i \frac{n_i m_i \tau_{\mathrm{v}}\left(\lambda(1-v_{z,i}/c);T_i\right)}{\rho_i} W_{\perp}(|\vec{r}_i|_{\perp}/h_i)  \textrm{.} \label{eq:sph-std-sigma}
\end{eqnarray}
The second is considerably more computationally intensive, and
interpolates each fluid quantity within the integrand {\it before}
performing the integration:
\begin{eqnarray}
\tau_{\mathrm{tot}}(\lambda) & = & \int_{-\infty}^{\infty} \dd z~ [n](\vec{r}) \tau_{\mathrm{v}} \left(\lambda\left( 1-[v_z](\vec{r})/c\right) ;[T](\vec{r}) \right) 
\end{eqnarray}
In the continuum limit $W\to \delta$ both these expressions reduce to
(\ref{eq:tau-sph}); further, it is simple to write a number of different
expressions midway between which have this same property.  For
quantities varying on scales $\gg h$ the two methods will give
identical results; however it is not clear a priori which, if any,
method is optimal for the DLA problem where as much information from
near the resolution limit needs to be extracted.

This problem is not confined to line of sight integrals, and so we may
turn for guidance to standard approaches to SPH. Although we have not
seen the ambiguity discussed explicitly, in general expressions are
used where the interpolation is performed as a final step as in
(\ref{eq:sph-std-sigma}). We therefore adopt this approach; however we
note that it has some odd properties, and in particular one may
define a dispersion in any quantity
\begin{eqnarray}
\sigma^2(r) \equiv [\vec{v}^2](\vec{r})-[\vec{v}]^2(\vec{r}) \ge 0
\end{eqnarray}
with equality holding if and only if $v(\vec{r})$ is constant within
the smoothing region. In other words, the standard method forces us to
accept a subresolution velocity dispersion in the fluid.  The exact
interpretation of this is somewhat elusive. Given a strictly
lagrangian interpretation, at any point there is a well defined
velocity and it should only vary over resolved spatial scales. But
owing to the method of interpolation, if we find nearby two SPH
particles on top of each other with equal and opposite velocities,
this would not lead to any significant spatial variation in
$v(\vec{r}) \simeq 0$ but would create a non-zero $[v]^2$. The
evolution of this ``turbulence'' would make for an interesting study,
not least to determine its physical meaning, but is beyond the scope
of this paper.

Our final expression for the line profile is therefore
\begin{equation}
\tau_{\mathrm{tot}}(\lambda) = \int_{-\infty}^{\infty} \dd z~ n_i \sigma\left(\lambda\left(1-[v_z]\right)/c;T_i, ([v^2]-[v]^2)(\vec{r}) \right)\label{eq:tau-sph-final}
\end{equation}
where the indicated thermal and turbulent broadening components are
added in quadrature. Because $\sigma^2$ is not a projected quantity,
the $z$ integration cannot be performed by simple integration of the
kernel, and therefore this quantity is performed as an explicit
numerical integration over the sightline.

Note that this intrinsic velocity dispersion should be distinguished
sharply from the line of sight velocity dispersion, e.g.
\begin{equation}
\langle v\rangle_{\himath}^2 - \langle v^2\rangle_{\himath} \equiv \left(\frac{\int \dd z [n_{\himath} v_{z}]}{\int \dd z [n_{\himath}]} \right) - \left(\frac{\int \dd z [n_{\himath} v^2_{z} ]}{\int \dd z [n_{\himath}]} \right) \label{eq:los-dispersion}
\end{equation}

It is clear, given the fiducial positioning of the SPH averaging
brackets $[]$ that this quantity will include contributions from the
intrinsic velocity dispersion, as well as genuine line of sight
dispersion effects.

\vspace{-0.5cm}
\section{Weighting Method as Discretization of an Integral}\label{sec:conv-meth-discr}

Let us consider a calculation of the line density of DLAs for the
situation described in Section \ref{sec:sightline-projection}, in
which $\sigma$ has some range of values with probability densities
$P(\sigma | M)$ for a given halo mass $M$. Our starting point is the
integral
\begin{equation}
  \frac{\dd N_{\mathrm{DLA}}}{\dd l} = \int \dd M \int \dd \sigma  \sigma f(M) P(\sigma | M)
\end{equation}
where we have temporarily adopted the physical distance measure $l$ to
obviate the need for $\dd l / \dd X$ on the right hand side.  We
discretize this equation by assigning all haloes to bins in virial
mass. Indexing these mass bins by $i$, and denoting the set of all
haloes within each mass bin by $\mathcal{H}_i$, one makes the
replacement $\int \dd M f(M) \ldots \to \sum_i F_i \ldots$ where $F_i$
is defined by eq.~(\ref{eq:F-binned}). 

Our probability distribution for $\sigma$, $P(\sigma | M)$, is
represented by individual haloes. The simplest way to represent $P$ is
therefore by a sum of $\delta$ functions\footnote{One could produce a
  more sophisticated method in which the $\delta$ functions are
  replaced by smooth functions with unit area to create a smoother
  final $P(\sigma | M)$.  However, it is not clear to what extent this
  is helpful and both methods reduce to the correct integral in the
  infinite-halo limit.} -- one for each halo -- normalized such that
the total probability $\int P(\sigma | M) \dd \sigma = 1$:
\begin{eqnarray}
  P(\sigma | M_i) & \to & \frac{1}{n(\mathcal{H}_i)}\sum_{h \in \mathcal{H}_i} \delta(\sigma_h-\sigma) \\
  \frac{\dd N_{\mathrm{DLA}}}{\dd l}&  \to  & \sum_i \frac{F_i}{n(\mathcal{H}_i)} \sum_{h \in \mathcal{H}_i} \sigma_h \label{eq:discrete-lden}
\end{eqnarray}
where $n(\cdots)$ counts the elements of its parameter set, and
$\sigma_h$ denotes the cross-section of the particular halo $h$.  This
expression reduces, as expected intuitively, to
\begin{equation}
\frac{\dd N_{\mathrm{DLA}}}{\dd l} = \sum_i N_i \bar{\sigma}_i
\end{equation}
where $\bar{\sigma}_i$ is the mean DLA cross-section of a halo in bin
$i$. 

We are next interested in computing, for some property $p$,

\begin{equation}
\frac{\dd^2 N_{\mathrm{DLA}}}{\dd l \dd p} = \int \dd \sigma \int \dd M \sigma n(M) P(\sigma\,\&\,p\, | M)  \textrm{.} \label{eq:continuuous-probabilistic-distrib}
\end{equation}
This will involve exactly the same replacements as before, with the only difference being
\begin{equation}
  P(\sigma\,\&\,p\, | M )\to  \sum_{h \in \mathcal{H}_i} \frac{\delta(\sigma_h-\sigma)}{n(\mathcal{H}_i)}
  \sum_{j \in \mathcal{S}_h}\frac{\delta(p_j-p)}{n(\mathcal{S}_h)}
  \textrm{.}\label{eq:double-P}
\end{equation} 
Here, $\mathcal{S}_h$ is the set of all DLA sightlines taken through
halo $h$. The expression (\ref{eq:double-P}) will leave us with a
floating $\delta$ function in the chosen property $p$, so the final
stage is to bin by this property. In other words, one integrates both
sides of the equation with respect to $p$ over a bin centred on $p_j$
with width $\Delta p$:
\begin{equation}
\left.\frac{\dd^2 N_{\mathrm{DLA}}}{\dd l\, \dd p}\right|_{p_j}  \Delta p = \sum_i \frac{F_i}{n(\mathcal{H}_i)} 
\sum_{h \in {H}_i} \sigma_h
\frac{n(\mathcal{P}_{j,h})}{n(\mathcal{S}_h)}
\end{equation}
in which we have introduced $\mathcal{P}_{j,h}$, the set of all
sightlines in halo $h$ such that $p$ falls within property bin $j$
($p_j- \Delta p/2 < p < p_j + \Delta p/2$). Note there is no sum over
sightlines in the above (only a counting of sightlines in particular
sets); the two sums are respectively over the mass bins and the haloes
within them. Since every halo falls into exactly one mass bin, this is
really just a sum over all haloes:
\begin{equation}
\left.\frac{\dd^2 N_{\mathrm{DLA}}}{\dd l\, \dd p}\right|_{p_j} = \sum_{h \in \mathcal{H}} \frac{N_{i(h)} \sigma_h}{n(\mathcal{H}_{i(h)})}  
\frac{n(\mathcal{P}_{j,h})}{n(\mathcal{S}_h) \Delta p} \textrm{.}\label{eq:discrete-p-final-binned}
\end{equation}
where $i(h)$ denotes the mass bin $i$ to which halo $h$ belongs.

For completeness, with all constants (including the conversion to
absorption distance $X$), our final calculation reads:

\begin{eqnarray}
  \left.\frac{\dd^2 N_{\mathrm{DLA}}}{\dd X\, \dd p}\right|_{p_j} & = & 
  \frac{c}{H_0 (1+z)^3} \sum_{h \in \mathcal{H}} \frac{w_h}{\Delta p}
  \frac{n(\mathcal{P}_{j,h})}{n(\mathcal{S}_h)} \textrm{;} \\
  w_h &  = & \frac{F_{i(h)} \sigma_h}{n(\mathcal{H}_{i(h)})}  \label{eq:wh} \textrm{.}
\end{eqnarray}

There is no essential barrier to investigating correlations between
properties by extending this method to produce $\dd ^3
N_{\mathrm{DLA}} / \dd X \dd p \dd q$ for some extra property
$q$. However, the growing sparsity of the data in the increasing
dimensional binning is such that an easier method suggests
itself. Correlations between parameters in DLA studies are generally
known only for $\mathcal{O}(100)$ observed systems. One can generate
from our simulations, using a Monte-Carlo technique, just such a
limited sample; this can be compared directly to observational data.
The probability for accepting a particular input sightline through
this technique must be proportional to the output line density
``generated'' by that individual sightline. From equation
(\ref{eq:discrete-lden}), it is clear that the contribution of halo
$h$ to the line-of-sight density of systems is given by a probability
$w_h$ (one may verify that this remains true for correlations between
any number of sightline properties) and thus the probability of
accepting that sightline should be proportional to this quantity.

\end{document}